\documentclass[12pt]{article}

\usepackage{amsmath}
\usepackage{amssymb}

\usepackage{correctmathalign} 

\usepackage{indentfirst}
\usepackage{amssymb}
\usepackage{amsfonts}
\usepackage{amscd}
\usepackage{amsbsy}
\usepackage{amsthm}
\usepackage{bm}
\usepackage{latexsym}
\usepackage{graphicx,color} 
\usepackage[dvipsnames]{xcolor}
\usepackage[colorlinks]{hyperref}
\hypersetup{linkcolor=blue,citecolor=blue,urlcolor=blue}

\usepackage[symbol]{footmisc} 

\def\nn{\nonumber}       
\def\beq{\begin{eqnarray}}
\def\eeq{\end{eqnarray}}

\def\al{\alpha}
\def\be{\beta}
\def\ch{\chi}
\def\ga{\gamma}
\def\de{\delta}

\def\ep{\epsilon}
\def\ze{\zeta}

\def\ka{\kappa}
\def\la{\lambda}

\def\pa{\partial}

\def\rh{\rho}
\def\si{\sigma}
\def\om{\omega}
\def\ph{\varphi}
\def\ta{\tau}
\def\th{\theta}

\def\La{\Lambda}
\def\Si{\Sigma}
\def\Om{\Omega}

\def\Th{\Theta}

\DeclareMathAlphabet\mathbfcal{OMS}{cmsy}{b}{n}

\usepackage[symbol]{footmisc} 
\usepackage{float} 
\usepackage{cite}  

\usepackage{orcidlink} 

\begin{document}

\begin{center}
\renewcommand*{\thefootnote}{\fnsymbol{footnote}} 
{\Large 
\bf SUSY QED with Lorentz-asymmetric fermionic matter and a glance at the electron's EDM}
\vskip 6mm
	
{\bf João Paulo S. Melo}
\orcidlink{0000-0001-5213-5183}
\hspace{-.5mm}\footnote{E-mail address: \ joaopaulo\_15@hotmail.com},
\,
{\bf Wagno Cesar e Silva}
\orcidlink{0000-0001-7832-3502}
\hspace{-.5mm}\footnote{E-mail address: \ wagnorion@gmail.com}
\,
and
\,
{\bf José A. Helayël-Neto}
\orcidlink{0000-0001-8310-518X}
\hspace{-.5mm}\footnote{E-mail address: \ helayel@cbpf.br}
\vskip 6mm

{
	Centro Brasileiro de Pesquisas Físicas,
	\\
	Rua Dr. Xavier Sigaud 150, Urca, 22290-180, Rio de Janeiro, RJ, Brazil
}

\end{center}
\vskip 2mm
\vskip 2mm


\begin{abstract}
	
\noindent
This contribution sets out to pursue the investigation of a supersymmetric electrodynamics model with Lorentz-symmetry violation (LSV) manifested by a space-time unbalance in the propagation of the fermionic charged matter. 
Despite violation of Lorentz symmetry, the supersymmetry algebra is kept untouched. A superspace approach is then adopted  to build up an $\mathcal{N}=1$-supersymmetric Abelian gauge theory in presence of a Lorentz-violating background supermultiplet that accommodates the space-time asymmetry parameter of the charged matter. It is described, in this scenario, how the particular Lorentz-symmetry breaking, brought about by the fermionic matter, affects its (matter) scalar partners and the photon/photino that minimally couple to charged matter. From the (modified) Dirac, Klein-Gordon and Maxwell field equations,  the corresponding dispersion relations are worked out to inspect and discuss the physical effects of the LSV Majorana fermion condensates that naturally emerge from the background supermultiplet. Finally, efforts are targeted to investigate the Gordon decomposition of the charged lepton electromagnetic current. This is carried out by iterating the (fermion and scalar) matter field equations, which points to an effective contribution to the electron's electric dipole moment. This result allows us to attain an estimate of the pseudo-vector condensate of the (LSV) Majorana background fermion. 
\vskip 3mm
	
	
\end{abstract}

\setcounter{footnote}{0} 
\renewcommand*{\thefootnote}{\arabic{footnote}} 
\section{Introduction}
\label{sec0}

Lorentz invariance stands as an essential cornerstone of Quantum Field Theory, manifesting as a fundamental symmetry within the framework of General Relativity and the Standard Model of Particle Physics.
However, in recent decades, the possibility that some spacetime symmetries could be violated at energies close to the Planck scale $(M_{\textrm{Planck}}\sim 10^{19}\,\textrm{GeV})$ has arisen in the context of different approaches, such as, e.g., String Theory \cite{KostSam89,KostPott91}, Loop Quantum Gravity \cite{GamPul99,Alf2000}
and Modified Electrodynamics \cite{CFJ90} (see also ref. \cite{Petrovbook} for a general review on Lorentz violation). At the same time, in this high-energy regime, Supersymmetry (SUSY) is expected to keep unbroken \cite{BgKost02}. In this way, it becomes pretty natural to wonder about the relationship between the Lorentz-symmetry violation (LSV) and SUSY breaking. Berger and Kostelecký were pioneers in proposing a method for incorporating Lorentz-breaking terms into a SUSY scenario; this can be accomplished by introducing the background anisotropies of LSV into the SUSY algebra \cite{BgKost02,Berger03}. Subsequently, a number of approaches to inspect supersymmetric extensions of Quantum Electrodynamics (QED) with LSV has emerged. These investigations have led to important phenomena such as vacuum birefringence \cite{Bolok05}, as well as the manifestation of SUSY breaking that arises from the mass splitting of superpartners \cite{KatSha06}. Alternatively to the original proposal presented by Berger and Kostelecký, the framework we adopt here to study LSV in a supersymmetric scenario consists in treating the Lorentz-breaking parameters as components of special (background) superfields, while SUSY algebra is kept unchanged. In this case, the central idea consists in seeking a supersymmetric formulation for the background tensor structures that constitute the spacetime anisotropies responsible for the LSV. This allows us to present a microscopic origin for the LSV parameters in terms of components of non-dynamical superfields, as it shall be subsequently described.

On the other hand, the possibility of LSV in other scenarios gives rise to interesting developments, e.g., in Condensed Matter Physics (CMP), where both SUSY and Lorentz invariance may appear as emergent symmetries.
The first indication of SUSY in CMP has come out from the study of a two-dimensional Ising model \cite{Fried84}, where an emergent SUSY becomes apparent near the critical point. In this context, the fundamental degrees of freedom may not exhibit SUSY, but an effective description via a supersymmetric action can be obtained from them. From this perspective, more recent works have contemplated scenarios of emergent SUSY in a $(2+1)$-dimensional system composed of a monolayer of carbon atoms arranged in a honeycomb lattice \cite{Ez08,Helayel11,DarCab13}. Moreover, such scenarios can also arise dynamically as pointed out, e.g., in refs. \cite{LeeSusy07,Grover14,PontLee14,Jian15}. It should also be mentioned that there are systems in low-energy CMP where the emergent Lorentz-symmetry plays an important role; it is the case of the so-called Dirac and Weyl materials \cite{Wehling14,Lv15,Tamai16,Yan17,Arnitage18,Gao19,Lee21}, topological insulators \cite{Qi08,Zhang09,Hasan10,Ryu10,Ando15,Xu16,Reja17} and graphene \cite{Novoselov04, Guinea09}. At the same time, in these scenarios, the violation of this emergent symmetry has been extensively explored in refs. \cite{Savrasov11, Grushin12, Goerbig16,  Bernevig15, Neves23}.
In this regard, the interesting work \cite{Kost22} presents a comprehensive exploration of quasi-particle excitations in the band structures of Dirac and Weyl materials. According to the authors, the emergent Lorentz-symmetry features in these materials shed light on Lorentz violation within the Standard Model Extension (SME) framework \cite{Coll97,Kost04,Kost11}.

Considering this perspective, the efforts of the present contribution are motivated by the theoretical study published by Isobe and Nagaosa of quantum critical phenomena in the phase transition between trivial and topological insulators in $(3+1)$ dimensions, which is described by a Dirac fermion coupled to the electromagnetic field \cite{IsoNag12}. In their model, LSV is realized by a constant parameter present only in the fermionic piece of the action. Our proposal consists in building up an N=1-supersymmetric $U(1)$ gauge model as an extended version of the Isobe-Nagaosa's paper with the LSV parameter no longer constant, but now taken to be space-time dependent. The main purpose is to understand how the LSV, initially manifested only in the fermionic sector, affects the sector of (charged) scalar partners in the gauge-invariant theory. For this, we shall adopt an alternative approach to implement LSV in connection with SUSY, as discussed in  refs. \cite{LCPT03,Helayel04}, in which LSV parameters are located in background superfields and they are therefore accompanied by background fermionic partners that condensate to give origin to bosonic LSV parameters.

In an interesting paper \cite{Shadmi06}, Katz and Shadmi contemplate the situation in which SUSY is broken together with Poincaré invariance, once a certain field acquires a Poincaré-violating ($x_\mu$-dependent) vacuum expectation value. This field yields a spontaneous (and, therefore, a soft) SUSY-breaking. A similar framework will be presented in details in the Subsection \ref{BCS} of this work. And we shall see, in the scenario we study, that a fermion, partner of the background scalar that violates Poincaré symmetry, can form (bilinear) condensates that, in turn, break Lorentz symmetry.

The outline of this paper is as follows: In Sect.~\ref{sec2}, we briefly review the extended Isobe-Nagaosa model, wherein the LSV parameter exhibits spacetime dependence, and present a formulation of the superspace consistent with such a model. In Sect.~\ref{sec3}, we work to get a supersymmetric $U(1)$ gauge model, based on the extended Isobe-Nagaosa model, and investigate the modifications induced by Lorentz-breaking background components. Next, in Section \ref{sec4}, we study the modified field equations and dispersion relations for both the fermionic/scalar matter sectors, and for the Maxwell sector as well. 
These results shall be used, in what follows, to obtain the rest energy and group velocity for some particular cases. In Sect. \ref{sec5}, we derive the effective electric dipole interaction and, based on current experimental results for the electron electric dipole moment (EDM), we present an estimate for one of the background Majorana fermion condensates. Finally, in Sect.~\ref{sec6}, we draw our Conclusive Considerations and Perspectives. 
Two Appendices supplement the main text. In the Appendix \ref{secA}, one can find an alternative approach to generate the $v_\mu\bar{\Psi}\ga^\mu \Psi$-term from a vector superfield. Next, in Appendix \ref{secB}, we collect
useful results concerning the calculation of our dispersion relations worked out in in Section \ref{sec4}.

In this work we use the signature $ (+,-,-,-) $ for
the Minkowski metric $\eta_{\mu\nu}$. Also, we adopt the natural system of units, with $c = 1$ and $\hbar = 1$.

\section{The extended Isobe-Nagaosa model}
\label{sec2}

We start off from the following extended version of the model proposed by Isobe-Nagaosa (see ref. \cite{JPSM24} for more details),
\begin{align}\label{Lagr_1}
\mathcal{L} 
= i \bar{\Psi}\Big[\ga^0 \pa_0 
-\textrm{Re}(\xi) \ga^i\pa_i\Big]\Psi
-\dfrac{1}{2}\Big\{\pa_i\textrm{Im}\big[\xi(x)\big]\Big\}\bar{\Psi}\ga^i\Psi
-m\bar{\Psi}\Psi,
\end{align}
where $\Psi$ is a Dirac fermion and $\xi$ is a factor that breaks Lorentz-symmetry. In the extended model \eqref{Lagr_1}, we introduce a space-time dependence on the background parameter $\xi$, since, in a general situation, such a parameter may not be constant and/or homogeneous, and it would be pertinent to exploit consequences of the non-constancy and non-homogeneity. In order to guarantee the reality of the action  associated with \eqref{Lagr_1}, it follows that the real part of $\xi$, $\textrm{Re}(\xi)$, must be a real constant parameter, while its imaginary part, $\textrm{Im}\big[ \xi (x)\big] \equiv I(x)$,  carries all the eventual space-time dependence of the $\xi$ factor. The original Isobe-Nagaosa model \cite{IsoNag12} is rescued considering $I(x)$ constant.

Before moving on, it is important to stress that, as it is explicit in the Isobe-Nagaosa's topological insulator QED model, the extended version we investigate in what follows \eqref{Lagr_1} also exhibits local U(1) invariance.

In what follows, it is convenient to introduce the covariant notation in the form
\begin{align}
\tilde{\pa}_\mu 
& = (\pa_0, \textrm{Re}(\xi) \pa_i), 
\label{4div} 
\\
v_\mu 
& = (0, -\pa_iI(x)).
\label{im_part} 
\end{align} 
Then, the Lagrangian \eqref{Lagr_1} can be brought into the form below:
\begin{align}
\mathcal{L}  
& = \dfrac{i}{2} \bar{\Psi}\ga^\mu(\tilde{\pa}_\mu \Psi)
-\dfrac{i}{2}(\tilde{\pa}_\mu\bar{\Psi}) \ga^\mu \Psi
+\dfrac{1}{2}v_\mu\bar{\Psi}\ga^\mu \Psi
-m\bar{\Psi}\Psi.  
\end{align}
As a step further, considering a Dirac spinor in the Weyl representation, one can arrange the $\Psi$-components according to their chirality,
\begin{align}\label{Dirac_spinor}
\Psi = 
\begin{pmatrix}
\psi_\al \\ \bar{\chi}^{\dot{\al}}
\end{pmatrix},  
\end{align}
where the Van der Waerden notation has been adopted. Defining $\si^\mu = (I_2, \si^i)$ and $\bar{\si}^\mu = (I_2, -\si^i)$, the $\ga$-matrices in Weyl representation are written as
\begin{align}
\gamma^\mu = 
\begin{pmatrix}
0 & \si^\mu \\
\bar{\si}^\mu & 0
\end{pmatrix} 
\;\;\;\;\;\textrm{and} \;\;\;\;\; 
\ga_5 = i \ga^0 \ga^1 \ga^2 \ga^3 = 
\begin{pmatrix}
-I_2 & 0 \\ 0 & I_2 
\end{pmatrix}.
\end{align}
Thus, the Lagrangian \eqref{Lagr_1} acquires the form
\begin{align}\label{Lagr_2}
\mathcal{L} = & 
\;\dfrac{i}{2}\left[\bar{\psi} \bar{\si}^\mu(  \tilde{\pa}_\mu\psi)-(\tilde{\pa}_\mu\bar{\psi}) \bar{\si}^\mu\psi+\bar{\chi} \bar{\si}^\mu (\tilde{\pa}_\mu\chi)-(\tilde{\pa}_\mu\bar{\chi})\bar{\si}^\mu\chi\right] 
\nn 
\\
& 
+\,\dfrac{1}{2}v_\mu(\bar{\psi}\bar{\si}^\mu\psi-\bar{\chi} \bar{\si}^\mu\chi)
-m(\bar{\psi}\bar{\chi}+\chi\psi),
\end{align}
with two chiral sectors, left and right, described by the $\psi$- and $\chi$-spinors, respectively.

\subsection{The SUSY-extended model in superspace}
\label{sec2-1}

In order to incorporate SUSY in the framework of the extended fermionic Lagrangian \eqref{Lagr_2}, and to carry out its coupling with the photon and photino, we adopt an approach that keeps SUSY algebra intact. We call into question that, instead of modifying SUSY algebra to include the LSV parameters, the latter are rather accommodated as components of a background superfield. To implement this point of view, one should remember the definition of four-divergence operator, $\tilde{\pa}_\mu$, presented in \eqref{4div} and define the position four-vector as follows:
\begin{align}
\tilde{x}_\mu=\bigg(x_0,-\dfrac{1}{ \textrm{Re}(\xi)}x_i\bigg),
\end{align}
in such a way that $\tilde{\pa}_\mu \tilde{x}^\nu = \de^\nu_\mu$. Then, we define the superspace coordinates, adopting the Salam-Strathdee formulation \cite{SS74}
(see, e.g., in ref. \cite{One83} for the textbook level introduction), as $\big\{\tilde{x}^\mu;\,\th_\al,\,\bar{\th}_{\dot{\al}}\big\}$ 
with $\mu$ being a space-time index and $\al=(1,2)$, as well as $\dot{\al}=(1,2)$, being indices of Grassmannian nature. That said, one can obtain the form of the SUSY generators,
\begin{align}\label{ger_Susy}
Q_\al = -i\pa_\al-(\si^\mu)_{\al\dot{\be}} \bar{\th}^{\dot{\be}} \tilde{\pa}_\mu,\;\;\;\;\;\bar{Q}_{\dot{\al}} = i \bar{\pa}_{\dot{\al}} + \th^{\be}(\si^\mu)_{\be\dot{\al}}\tilde{\pa}_\mu,
\end{align}
and establish the anti-commutation relations
\begin{align}
\left\{D_\al, Q_\be\right\}=\left\{ \bar{D}_{\dot{\al}}, \bar{Q}_{\dot{\be}} \right\}=\left\{D_\al,\bar{Q}_{\dot{\be}} \right\} = \left\{\bar{D}_{\dot{\al}}, Q_\be\right\}=0,
\end{align}
with the SUSY covariant derivative operators given by
\begin{align}
D_\al & = \pa_\al+i(\si^\mu)_{\al\dot{\be}} \bar{\th}^{\dot{\be}}\tilde{\pa}_\mu, \\
\bar{D}_{\dot{\al}} & = \bar{\pa}_{\dot{\al}}+i\th^\be (\si^\mu)_{\be\dot{\al}}\tilde{\pa}_\mu.
\end{align}
From \eqref{ger_Susy}, it is clear that the SUSY algebra is left untouched, as already mentioned.

\section{$\mathcal{N}=1$-supersymmetric $U(1)$ gauge-field sector}
\label{sec3}

In what follows, we shall apply the ideas presented in the previous Section to build up a supersymmetric version of QED.  First, let us investigate the modifications the formulation of superspace, based on the fermionic matter sector, in  the gauge field sector.

\subsection{The Maxwell sector}
\label{sec3-1}

Consider a real vector superfield, $A(\tilde{x},\th,\bar{\th})$, in the Wess-Zumino gauge 
(see, e.g., the textbook in ref. \cite{One83} for more details),
\begin{align} \label{WZG}
A_{\textrm{WZ}}=i\th^{2}\bar{\th}\bar{\la}-i\bar{\th}^{2}\th\la+(\th\si^\mu\bar{\th})\tilde{a}_\mu+\dfrac{1}{2}\th^{2}\bar{\th}^{2}K, 
\end{align}
where $\la$ is the Weyl component of a Majorana spinor, $\tilde{a}_\mu$ is the four-vector and $K$ is an auxiliary field. To be consistent with the definitions of superspace and superfields we are considering, $\tilde{a}_\mu$ must have the form
\begin{align}\label{4vec}
\tilde{a}_\mu=\left(A_0,-\frac{1}{\textrm{Re}(\xi)}A_i\right).
\end{align} 
Here, it is clear that the Lorentz violation in the Maxwell sector is signaled exclusively by the presence of $\textrm{Re}(\xi)$. Hence, we can expect that this sector is not affected by the space-time dependence of the background.

In this case, the generalization of the Abelian gauge condition for the modified electromagnetic four-potential is given by
\begin{align} \label{abelG}
\tilde{a}_\mu \rightarrow \tilde{a}'_\mu = \tilde{a}_\mu-\tilde{\pa}_\mu\big[2\textrm{Im}(z)\big],
\end{align}
with $\textrm{Im}(z)$ standing for the imaginary part of the complex scalar component of some chiral superfield.
The kinetic term for $\tilde{a}_\mu$ can be generated by the contractions
\begin{align}
W^\al W_\al\Big\vert_{\sim\th^{2}} & = -\dfrac{1}{2} \mathcal{F}^{\mu\nu}\mathcal{F}_{\mu\nu}+\dfrac{i}{2} {^{*\!\!}\mathcal{F}^{\mu\nu}}\mathcal{F}_{\mu\nu }-2i\la\si^\mu\tilde{\pa}_{\mu}\bar{\la}+K^2, \label{WW1} \\
\bar{W}_{\dot{\al}}\bar{W}^{\dot{\al}}\Big\vert_{\sim\bar{\th}^{2}} & = -\dfrac{1}{2}\mathcal{F}^{\mu\nu}\mathcal{F}_{\mu\nu}-\dfrac{i}{2}{^{*\!\!}\mathcal{F}^{\mu\nu}}\mathcal{F}_{\mu\nu}+2i(\tilde{\pa}_{\mu}\la)\si^\mu\bar{\la}+(K^*)^2, \label{WW2}
\end{align}
where $\mathcal{F}_{\mu\nu}$ is the electromagnetic field strength and $\la$ is identified as the photino spinor field. In the expressions \eqref{WW1} and \eqref{WW2}, we define the spinorial superfield, $ W_\al = -\dfrac{1}{4}\bar{D}\bar{D}D_\al A_{\textrm{WS}} $, with $\bar{W}_{\dot{\alpha}}$ representing its complex conjugate, and use the compact notations
\begin{align}\label{FF}
\mathcal{F}_{\mu\nu}=(\tilde{\pa}_\mu\tilde{a}_\nu-\tilde{\pa}_\nu\tilde{a}_\mu),\;\;\;\;\; {^{*\!\!}\mathcal{F}^{\mu\nu}}=\dfrac{1}{2}   \varepsilon^{\mu\nu\ka\rho}\mathcal{F}_{\ka\rho}.
\end{align}
With all described above, the supersymmetric gauge field action reads as follows: 
\begin{align}\label{SMaxwell}
S_\textrm{M} & = \dfrac14\int d^4 x\,\biggl\{\int d^2 \th\, W^{2}+\int d^2\bar{\th}\,\bar{W}^{2}\biggr\}
\end{align}
or, in terms of the component fields,
\begin{align}
S_\textrm{M} & = \frac{1}{2}\int d^4x\bigg\{-\dfrac{1}{2}\mathcal{F}^{2}_{\mu\nu}+i\Big[(\tilde{\pa}_{\mu}\bar{\la})\bar{\si}^\mu\la-\bar{\la}\bar{\si}^\mu\tilde{\pa}_{\mu}\la\Big]+|K|^2\bigg\}.
\end{align}
In Equation\,\eqref{SMaxwell} we use the compact notations  $W^2$ and $\bar W^2$ for the definitions \eqref{WW1} and \eqref{WW2}, respectively.
It is important to notice that, in our case, the modified Maxwell sector is obtained from the matter sector as a consequence of SUSY. In Section \ref{sec4}, we shall explicitly see the modifications in Maxwell's equations due to the prescription \eqref{4vec}.

\subsection{The matter sector}
\label{sec3-2}

Our next task is to supersymmetrize the matter sector based on the Lagrangian \eqref{Lagr_2}. This procedure will lead to the kinetic and mass terms for the scalar supersymmetric partners of $\psi$ and $\chi$.

To have a supersymmetric version of the left sector in \eqref{Lagr_2}, we define a chiral superfield,
\begin{align}
\Phi_\textrm{L} & = e^{i(\th\si^\mu\bar{\th})\tilde{\pa}_\mu} \left(\ph+\sqrt{2}\th\psi+\th^{2}f\right) \nn \\
& =\ph+\sqrt{2} \th\psi+\th^{2}f+i(\th\si^\mu\bar{\th})\tilde{\pa}_\mu\ph-\frac{i}{\sqrt{2}}\th^{2}(\tilde{\pa}_\mu\psi)\si^\mu\bar{\th}-\dfrac{1}{4}\th^{2}\bar{\th}^{2}\widetilde{\square}\ph,
\end{align}
and its  anti-chiral associated superfield,
\begin{align}
\bar{\Phi}_\textrm{L} & = \ph^*+\sqrt{2}\bar{\th}\bar{\psi}+\bar{\th}^{2}f^*-i(\th\si^\mu\bar{\th})\tilde{\pa}_\mu\ph^*+\frac{i}{\sqrt{2}}\bar{\th}^{2}\th\si^\mu\tilde{\pa}_\mu\bar{\psi}-\dfrac{1}{4}\th^{2}\bar{\th}^{2}\widetilde{\square}\ph^*,
\end{align}
where $\ph$ is a charged scalar field, $\psi$ is the Weyl component of a Dirac spinor and $f$ is an auxiliary field. 
Similarly, for the right sector, we have
\begin{align}
\bar{\Phi}_{\textrm{R}} & = s+\sqrt{2}\bar{\th}\bar{\chi}+\bar{\th}^{2}g-i(\th\si^\mu\bar{\th})\tilde{\pa}_{\mu}s+\frac{i}{\sqrt{2}}\bar{\th}^{2}\th\si^\mu\tilde{\pa}_\mu\bar{\chi}-\dfrac{1}{4}\th^{2}\bar{\th}^{2}\widetilde{\square}s, \\
\Phi_{\textrm{R}} & = s^*+\sqrt{2}\th\chi+\th^{2}g^*+i(\th\si^\mu\bar{\th})\tilde{\pa}_\mu s^*-\frac{i}{\sqrt{2}}\th^{2}(\tilde{\pa}_\mu\chi)\si^\mu\bar{\th}-\dfrac{1}{4}\th^{2}\bar{\th}^{2}\widetilde{\square}s^*.
\end{align}
Here, $s$, $\ch$ and $g$ are the right component fields analogous to $\ph$, $\psi$ and $f$ in the left sector, respectively.  It is important to notice that the introduction of two chiral superfields with independent fermionic components, $\psi$ and $\ch$, is necessary in order to compose the (charged) Dirac fermion $\Psi$. 
The mass terms arise from the combinations
\begin{equation}
\Phi_{\textrm{R}}\Phi_{\textrm{L}}\Big\vert_{\sim\th^{2}}=s^*f +g^*\ph-\chi\psi\;\;\;\textrm{and}\;\;\;\bar{\Phi}_{\textrm{L}} \bar{\Phi}_{\textrm{R}}\Big\vert_{\sim\bar{\th}^{2}} =f^*s+\ph^*g-\bar{\psi}\bar{\chi}.
\end{equation}

To implement the Abelian gauge symmetry in the matter sector, we consider the transformation of chiral superfields under the global $ U(1) $ gauge transformation by the relations
\begin{align}
& \Phi_{\textrm{L}}^{\prime} = e^{ie \La} \Phi_{\textrm{L}},\;\;\;\;\bar{\Phi}_{\textrm{L}}^{\prime}=e^{-ie \bar{\La}}\bar{\Phi}_{\textrm{L}}, \label{t1} \\
& \bar{\Phi}_{\textrm{R}}^{\prime} = e^{ie \bar{\La}}\bar{\Phi}_{\textrm{R}},\;\;\;\;\Phi_{\textrm{R}}^{\prime}=e^{-ie\La}\Phi_{\textrm{R}}, \label{t2}
\end{align}
where $ e $ is the $ U(1) $ charge of the superfield $ \Phi $. Let us remark that the mass term in the supersymmetric action is automatically gauge invariant. 

After some algebra, one can verify that the kinetics terms in \eqref{Lagr_2} are generated by combinations like $ \big(\Phi_{\textrm{L}}e^{eA_{\textrm{WZ}}}\bar{\Phi}_\textrm{L}+\Phi_{\textrm{R}}e^{-eA_{\textrm{WZ}}}\bar{\Phi}_\textrm{R}\big) \big\vert_{\sim\th^{2} \bar{\th}^{2}} $.

\subsection{The $v_\mu\bar{\Psi}\ga^\mu \Psi$-term}
\label{sec3-3}

The less immediate structure to supersymmetrize is the term
\begin{align}\label{cond1}
\dfrac{1}{2} v_\mu  \left( \bar{\psi} \bar{\si}^\mu \psi -\bar{\chi} \bar{\si}^\mu \chi \right).
\end{align}
The central point of our approach is to consider LSV in a supersymmetric scenario. The background supermultiplet which accommodates the Lorentz-symmetry violating parameters realizes the (explicit) breaking of Lorentz symmetry and, contemporarily, induces spontaneous SUSY breaking \cite{Belich13,Bonetti18,Terin22}. 

We have actually three alternatives to accomplish this program: $ 1) $ to introduce an extra chiral-superfield and then identify $v_\mu$ as four-gradient of a (background) scalar; $ 2) $ to define a vector superfield and consider $v_\mu$ as a genuine four-vector; $ 3) $ to introduce the LSV four-vector $v_\mu$ by means of an explicit breaking of SUSY, so that not only Lorentz symmetry but also SUSY be also explicitly broken. In this case, the background four-vector would not have background fermionic partners. The latter give rise to bilinears that may be identified with background vector or tensors. Since we are interested in considering this situation, we are going to discard path (3).

From a conceptual point of view, we can consider that the anisotropy arises from some kind of more fundamental field that may condensate to form the LSV background. As we argue, it is important to look for a microscopic origin for the anisotropy and, in this case, the first approach becomes more natural, since the background field does not appear by itself; contrary, it is derived from some more fundamental object; in our case, a scalar, as we shall soon see. Let us remember that, in general, it is the scalars that trigger transitions between different phases of the primordial Universe. Hence, they may have played an important role in the formation of anisotropic structures during the evolution of the Universe. These are motivations that support our choice of the background as a chiral superfield and not a four-vector supermultiplet.

In what follows, we shall adopt the first scenario. For completeness, though we are not going to adopt the alternative (2), we cast in Appendix \ref{secA} the results obtained from the alternative approach, i.e., the case of a vector superfield. 

\subsubsection{Background chiral superfield} \label{BCS}

Let us introduce an additional chiral superfield $\Om$, which is gauge invariant from the perspective of the transformation in \eqref{t1} and \eqref{t2}. The $\th$-expansion of $\Om$ reads
\begin{align}\label{background}
\Om & = \om+\sqrt{2} \th\ze+\th^{2}h+i(\th\si^\mu\bar{\th})\tilde{\pa}_\mu \om-\frac{i}{\sqrt{2}}\th^{2}(\tilde{\pa}_\mu\ze)\si^\mu \bar{\th}-\dfrac{1}{4}\theta^{2}\bar{\th}^{2}\widetilde{\square}\om,
\end{align}
where $\om$ is a scalar field, $\ze$ is the Weyl component of a Majorana spinor and $h$ is an auxiliary field. One should also notice that, differently from the physical propagating superfields, the background superfield $\Om$ has trivial canonical dimension. 
The products of superfields which can reproduce \eqref{cond1} are
\begin{align}
(\Om+\bar{\Om})\Phi_{\textrm{L}}e^{eA_{\textrm{WZ}}}\bar{\Phi}_\textrm{L} \Big\vert_{\sim\th^{2} \bar{\th}^{2}} = & 
-\frac{i}{2}\tilde{\pa}_\mu(\om-\om^{*})\bar{\psi}\bar{\si}^\mu\psi+(\om+\om^{*})\bigg\{\frac{i}{2}\Big[(\tilde{\pa}_\mu\bar{\psi})\bar{\si}^\mu\psi-\bar{\psi}\bar{\si}^\mu\tilde{\pa}_\mu\psi\Big] 
\nn 
\\
& -\dfrac{1}{2}\Big[(\widetilde{\square}\ph^*)\ph+\ph^*\widetilde{\square}\ph\Big]+f^*f\bigg\}+f^*(h\ph-\ze\psi)+f(h^*\ph^*-\bar{\ze}\bar{\psi}) 
\nn 
\\
& +\dfrac{1}{2}\tilde{\pa}_\mu(\om-\om^{*})\Big[(\tilde{\pa}^\mu\varphi^*)\ph-\ph^*\tilde{\pa}^\mu\ph\Big]+\frac{i}{2}\Big[\ph(\tilde{\pa}_\mu\bar{\psi})\bar{\si}^{\mu}\ze-\ph\bar{\psi}\bar{\si}^{\mu}\tilde{\pa}_\mu\ze 
\nn 
\\
& -\bar{\psi}\bar{\si}^{\mu}\ze\tilde{\pa}_\mu\ph\Big]-\frac{i}{2}\Big[\ph^*\bar{\ze}\bar{\si}^{\mu}\tilde{\pa}_\mu\psi-\ph^*(\tilde{\pa}_\mu\bar{\ze})\bar{\si}^{\mu}\psi-\bar{\ze}\bar{\si}^{\mu}\psi\tilde{\pa}_\mu\ph^*\Big] 
\nn 
\\
& -\frac{e(\om+\om^{*})}{2}\bigg\{i\sqrt{2}(\bar{\la}\bar{\psi}\ph-\la\psi\ph^{*})-i(\ph^{*}\tilde{\pa}^{\mu}\ph-\ph\tilde{\pa}^{\mu}\ph^{*})\tilde{a}_{\mu} 
\nn 
\\
& +(\bar{\psi}\bar{\si}^\mu\psi)\tilde{a}_{\mu}-\bigg(K+\frac{e}{2}\tilde{a}^{2}\bigg)\ph^{*}\ph\bigg\}+\frac{e}{2}\bigg\{i\tilde{\pa}^{\mu}(\om-\om^{*})\,\ph^{*}\ph\,\tilde{a}_{\mu} 
\nn 
\\
& -\big[(\bar{\psi}\bar{\si}^{\mu}\ze)\ph+(\bar{\ze}\bar{\si}^{\mu}\psi)\ph^{*}\big]\tilde{a}_{\mu}-i\sqrt{2}(\bar{\la}\bar{\ze}-\la\ze)\ph^{*}\ph\bigg\},
\end{align}
and an analog structure working to reproduce the right sector.

Choosing the scalar component $\om$ of the background supermultiplet $\Om$ as it is done in ref. \cite{LCPT03},
\begin{equation}\label{w_conditions}
(\om+\om^{*})=0 \;\;\;\textrm{and}\;\;\; (\om-\om^{*})=2i\varpi,
\end{equation}
with $ \varpi=v_\nu\tilde{x}^{\nu} $, we must impose that $ \tilde{\pa}_\mu v_{\nu}=0 $, i.e., that the imaginary part of parameter $ \xi $ in Equation\,\eqref{im_part} has an only linear dependence on the coordinates. 

At this point, we again call into question the paper by Katz and Shadmi \cite{Shadmi06}, where an $x_\mu$-dependent scalar field violates Poincaré symmetry and, as a consequence, softly  breaks SUSY. This is exactly what is happening above, where the imaginary part of the $\omega$-background field is taken linearly dependent on $\tilde{x}_\mu$. Actually, this field plays the role of the spurion considered in ref. \cite{Shadmi06}.  In due time, we should mention the paper by Nibbelink and Pospelov \cite{Pospelov05}, where the authors, classify LSV operators compatible with exact SUSY, restricting their analysis to vector and tensor backgrounds, and show that LSV operators in the context of the MSSM must have at least dimension five; they are therefore Planck mass-suppressed. This is not what we are doing here, since we are not based on exact SUSY; contrary, we associate a soft (spontaneous) SUSY breaking with Poincaré symmetry violation. We then justify that we have LSV operators with dimensions less than five because we are not bound to considering exact SUSY. As already anticipated in the Introduction, the $\zeta$-fermion given above, partner of the spurion in the $\Omega$-superfield, will be responsible for condensates that realize LSV contributions to the matter action. 

From these considerations, we can write our supersymmetric model as follows below:
\begin{align}\label{sA1}
S = & 
\,\int d^{4}x\,\bigg\{\int d^{2}\th d^{2}\bar{\th}\,\bigg[\bar{\Phi}_{\textrm{L}}e^{eA_{\textrm{WZ}}}\Phi_{\textrm{L}}+\bar{\Phi}_{\textrm{R}}e^{-eA_{\textrm{WZ}}}\Phi_{\textrm{R}}+\frac{1}{2}(\Om+\bar{\Om})\big(\bar{\Phi}_{\textrm{L}}e^{eA_{\textrm{WZ}}}\Phi_{\textrm{L}} 
\nn 
\\
& 
-\,\bar{\Phi}_{\textrm{R}}e^{-eA_{\textrm{WZ}}}\Phi_{\textrm{R}}\big)\bigg]+\frac14\bigg(\int d^2 \th\, W^{2}+c.c.\bigg)+m\bigg(\int d^{2}\th\,\Phi_{\textrm{R}}\Phi_{\textrm{L}}+c.c.\bigg)\bigg\}.
\end{align}
Using the field equations for the auxiliary fields, \eqref{sA1} can be formulated in terms of the physical fields:

\begin{align}\label{sA1_onShp}
S_{1} = & 
\,\int d^{4}x\,\bigg\{-\frac14\mathcal{F}^{2}_{\mu\nu}+\frac{i}{2}\Big[(\tilde{\pa}_\mu\bar{\la})\bar{\si}^{\mu}\la-\bar{\la}\bar{\si}^{\mu}\tilde{\pa}_\mu\la\Big]-\dfrac{i}{2}\Big[\bar{\psi}\bar{\si}^\mu  \tilde{\pa}_\mu\psi-(\tilde{\pa}_\mu\bar{\psi}) \bar{\si}^\mu\psi+\bar{\chi} \bar{\si}^\mu \tilde{\pa}_\mu\chi 
\nn 
\\
& 
-\,(\tilde{\pa}_\mu\bar{\chi})\bar{\si}^\mu\chi\Big]+\dfrac{1}{2}v_\mu(\bar{\psi}\bar{\si}^\mu\psi-\bar{\chi}\bar{\si}^\mu\chi)-m(\bar{\psi}\bar{\chi}+\chi\psi)+\Big[(\tilde{\pa}^\mu\ph^*)\tilde{\pa}_\mu\ph+(\tilde{\pa}^\mu s^*)\tilde{\pa}_\mu s\Big] 
\nn 
\\
&
+\,\dfrac{i}{2}v_\mu\Big[(\tilde{\pa}^\mu\varphi^*)\ph-\ph^*\tilde{\pa}^\mu\ph+(\tilde{\pa}^\mu s^*)s-s^*\tilde{\pa}^\mu s\Big]-\bigg(m^{2}+\frac{1}{4}h^{*}h\bigg)(\ph^{*}\ph+s^{*}s) 
\nn 
\\
& 
+\,\frac{i}{2}\Big[\ph(\tilde{\pa}_\mu\bar{\psi})\bar{\si}^{\mu}\ze-\ph^*\bar{\ze}\bar{\si}^{\mu}\tilde{\pa}_\mu\psi+s\bar{\ze}\bar{\si}^{\mu}\tilde{\pa}_\mu\chi-s^{*}(\tilde{\pa}_\mu\bar{\chi})\bar{\si}^{\mu}\ze\Big]-\frac{1}{2}\bigg[\ph\bigg(m\,\ze\ch-\frac{1}{2}h\bar{\ze}\bar{\psi}\bigg) 
\nn 
\\
& 
+\,\ph^{*}\bigg(m\,\bar{\ze}\bar{\ch}-\frac{1}{2}h^{*}\ze\psi\bigg)-s\bigg(m\,\bar{\ze}\bar{\psi}+\frac{1}{2}h^{*}\ze\ch\bigg)-s^{*}\bigg(m\,\ze\psi+\frac{1}{2}h\bar{\ze}\bar{\ch}\bigg)\bigg] 
\nn 
\\
& 
-\,\frac{1}{4}\Big(\bar{\ze}\bar{\psi}\ze\psi+\bar{\ze}\bar{\chi}\ze\chi\Big)-\frac{e}{2}\Big[i\sqrt{2}(\ph\bar{\la}\bar{\psi}-\ph^{*}\la\psi+s\la\ch-s^{*}\bar{\la}\bar{\ch})+(\bar{\psi}\bar{\si}^{\mu}\psi-\bar{\ch}\bar{\si}^{\mu}\ch)\tilde{a}_{\mu} 
\nn 
\\
&
+\,i(\ph\,\tilde{\pa}^{\mu}\ph^{*}-\ph^{*}\tilde{\pa}^{\mu}\ph+s\,\tilde{\pa}^{\mu}s^{*}-s^{*}\tilde{\pa}^{\mu}s)\tilde{a}_{\mu}\Big]-\frac{e}{4}\Big[(\bar{\psi}\bar{\si}^{\mu}\ze)\ph+(\bar{\ze}\bar{\si}^{\mu}\psi)\ph^{*}+(\bar{\ch}\bar{\si}^{\mu}\ze)s^{*} 
\nn 
\\
&
+\,(\bar{\ze}\bar{\si}^{\mu}\ch)s\Big]\tilde{a}_{\mu}-\frac{e}{2}(\ph^{*}\ph+s^{*}s)\bigg[v^{\mu}\tilde{a}_{\mu}+\frac{i}{\sqrt{2}}(\bar{\la}\bar{\ze}-\ze\la)-\frac{e}{2}\tilde{a}^{2}\bigg]
\nn 
\\
&
-\,\frac{e^{2}}{8}(\ph^{*}\ph-s^{*}s)^{2}\bigg\}.
\end{align}
Despite the lengthy expression in $S_1$, it is immediate to identify the fermionic model \eqref{Lagr_1} in the first two lines. In order to speed up the analysis of the contributions arising from the presence of the background, we group the terms according to sectors. 
It proves much more useful to rewrite the result \eqref{sA1_onShp} in terms of 4-component spinors. 

In the gauge sector, we have the following standard structure
\begin{align}\label{s_calibre}
S_{\textrm{gauge}} & = -\frac14\int d^{4}x\,\bigg\{\mathcal{F}^{2}_{\mu\nu}-i\Big[(\tilde{\pa}_\mu\bar{\La})\ga^{\mu}\La-\bar{\La}\ga^{\mu}\tilde{\pa}_\mu\La\Big]\bigg\},
\end{align}	
where the gaugino (photino) is a Majorana fermion written as
\begin{align}\label{fotino}
\La = 
\begin{pmatrix}
\la_\al \\ \bar{\la}^{\dot{\al}}
\end{pmatrix}.
\end{align}
It should be recalled that that the background component-fields are not present in the photon/photino sector, in contrast to other scenarios with LSV, such as, e.g., the supersymmetric extensions of the Carroll-Field-Jackiw model of refs. \cite{LCPT03,belich15}. The reason for this is that, in our case, the gauge sector is affected only by the real part of the background parameter $\xi$. In fact, the modifications in the gauge sector, induced by the background, only manifest as ``dilations" in Maxwell's equations, as we shall see in the next Section.

Using \eqref{Dirac_spinor} and defining the covariant derivative $ \tilde{D}_\mu\equiv\tilde{\pa}_\mu-\dfrac{ie}{2}\tilde{a}_{\mu} $, we can organize the fermionic sector, including the gauge coupling field, in the form
\begin{align}\label{s_fermiônico}
S_{\textrm{fermion}} = & 
\,\int d^{4}x\,\bigg\{\dfrac{i}{2}\Big[(\tilde{D}^{*}_\mu\bar{\Psi})\ga^\mu\Psi-\bar{\Psi}\ga^\mu  \tilde{D}_\mu\Psi\Big]+\dfrac{1}{2}v_\mu(\bar{\Psi}\ga^\mu\Psi)-m\bar{\Psi}\Psi 
\nn 
\\
& -\frac{1}{16}(\bar{Z}\ga_{\mu}\ga_{5}Z)(\bar{\Psi}\ga^{\mu}\ga_{5}\Psi)\bigg\},
\end{align}
with the Majorana spinor $ Z $ (the background fermion), given by
\begin{align}
Z = 
\begin{pmatrix}
\ze_\al \\ \bar{\ze}^{\dot{\al}}
\end{pmatrix}.
\end{align}
The term with dependence on the background fermionic condensate $\bar{Z}\ga_{\mu}\ga_{5}Z$ that appears in the fermionic sector is an additional contribution to the mass of the Dirac spinor $\Psi$. 
In Equation\,\eqref{background}, in defining the $\Om$-superfield, the component fields
$\om,\,\ze$ and $h$ are space-time dependent. However, since $\Om$ is non-dynamical (it is a background superfield), we are allowed to
fix its components. Indeed, in Equation\,\eqref{w_conditions}, by taking $\textrm{Re}(\om) = 0$  and
$\textrm{Im}(\om) = v^\mu \tilde{x}_\mu$, $\ze$ and $h$ are taken to be constant component fields,
and this choice is compatible with the SUSY transformations 
of the component fields. Therefore, the $Z$-bilinear is constant, so that
it is legitimate to state that it is a contribution to the mass-like 
$\bar{\Psi}\ga^\mu \ga_5 \Psi$-term.
It is interesting to notice that this correction can be identified as the coefficient $b_\mu$ in one of the CPT-odd terms introduced in the SME \cite{Coll97} in order to implement the violation of the Lorentz symmetry.
  
The (gauge-invariant) sector of matter scalar partners reads as given below:
\begin{align}\label{s_escalar}
S_{\textrm{scalar}} = & 
\,\int d^{4}x\,\bigg\{(\tilde{D}^\mu\ph^*)\tilde{D}_\mu\ph+(\tilde{D}^\mu s^*)\tilde{D}_\mu s-iv_\mu\big(\ph^*\tilde{D}^\mu\ph+s^*\tilde{D}^\mu s\big) 
\nn 
\\
& 
-\bigg(m^{2}+\frac{1}{4}h^{*}h\bigg)(\ph^{*}\ph+s^{*}s)\bigg\}.
\end{align}
As in the fermionic sector, there is a correction to the scalar masses, $\ph$ and $s$; in this case, proportional to $h^{*}h$. This non-degeneracy in the mass spectrum indicates a SUSY breaking due to the presence of the background. Furthermore, there is also a contribution of the background vector, $v_\mu$, to the dynamic part of the scalar sector. On the other hand, in the fermionic case, $v_\mu$ only affects the mass term, as we can see in the first line of \eqref{s_fermiônico}.

Finally, we have the gaugino and matter self-interaction contributions to the Lagrangian:
\begin{align}\label{s_interação}
S_{\textrm{mixing$+$int}} = & 
-\int d^{4}x\,\bigg\{\frac{ie}{\sqrt{2}}\Big[\bar{\Psi}(\ph\mathcal{P}_{\textrm{R}}+s\mathcal{P}_{\textrm{L}})\La-\bar{\La}(\ph^{*}\mathcal{P}_{\textrm{L}}+s^{*}\mathcal{P}_{\textrm{R}})\Psi\Big] 
\nn 
\\
& 
+\,\frac{i}{2}\Big[\bar{Z}\ga^{\mu}(\ph^*\mathcal{P}_{\textrm{L}}-s^{*}\mathcal{P}_{\textrm{R}})\tilde{D}_\mu\Psi
-(\tilde{D}^{*}_\mu\bar{\Psi})\ga^{\mu}(\ph\mathcal{P}_{\textrm{L}}-s\mathcal{P}_{\textrm{R}})Z\Big]
\nn 
\\
& 
+\,\frac{1}{2}\bar{\Psi}\bigg[\bigg(m\ph-\frac{1}{2}h^{*}s\bigg)\mathcal{P}_{\textrm{L}}-\bigg(ms+\frac{1}{2}h\ph\bigg)\mathcal{P}_{\textrm{R}}\bigg]Z \nn \\
& +\,\frac{1}{2}\bar{Z}\bigg[\bigg(m\ph^{*}-\frac{1}{2}hs^{*}\bigg)\mathcal{P}_{\textrm{R}}-\bigg(ms^{*}+\frac{1}{2}h^{*}\ph^{*}\bigg)\mathcal{P}_{\textrm{L}}\bigg]\Psi 
\nn 
\\
& 
+\,\frac{e}{2}\bigg[\frac{i}{2\sqrt{2}}(\bar{\La}\ga_{5}Z+\bar{Z}\ga_{5}\La)\bigg](\ph^{*}\ph+s^{*}s)+\frac{e^{2}}{8}(\ph^{*}\ph-s^{*}s)^{2}\bigg\},
\end{align}
with the parity operators,
\begin{align}
\mathcal{P}_{\textrm{R}} =\frac{I_{4}+\ga_{5}}{2}= \begin{pmatrix}
0 & 0 \\ 0 & I_{2}
\end{pmatrix}\;\;\;\;\;\textrm{and}\;\;\;\;\;\mathcal{P}_{\textrm{L}} =\frac{I_{4}-\ga_{5}}{2}= 
\begin{pmatrix}
I_{2} & 0 \\ 0 & 0
\end{pmatrix}.
\end{align}
It is worth mentioning that the new couplings induced by the background fermion, $Z$, except for those involving the degrees of freedom of the gauge sector ($a_\mu$ and $\La$), must be taken into account when calculating the dispersion relations for the matter fields, because such couplings do not characterize interaction terms (since the background is not dynamic).

\section{Field equations and dispersion relations}
\label{sec4}

With the supersymmetric action \eqref{sA1_onShp} in hand, we proceed, in this Section, to a detailed analysis of the field equations in both the matter and gauge sectors. Subsequently, in each case, we shall derive the modified dispersion relations within this framework, aiming to understand the implications of LSV in connection with SUSY.

\subsection{The modified Maxwell's equations}
\label{sec4-1}

The modified Maxwell's equations in the vacuum come from the $\mathcal{F}^{2}_{\mu\nu}$ part of the action \eqref{s_calibre},
\begin{align}\label{MXeq}
\tilde{\pa}_\mu \mathcal{F}^{\mu\nu} = 0.
\end{align}
Let us recall that $\mathcal{F}_{\mu\nu} = (\tilde{\pa}_\mu \tilde{a}_\nu - \tilde{\pa}_\nu \tilde{a}_\mu )$. From \eqref{MXeq} we obtain the modified version of the Gauss's law for the electric field and the Ampère-Maxwell equation, in the simplified form,
\begin{align}
& \textrm{Re}(\xi)\nabla\cdot\mathbfcal{E} = 0, \label{eqMax1} \\
& \textrm{Re}(\xi) \nabla\times\mathbfcal{B}=\pa_t\mathbfcal{E}, \label{eqMax2}
\end{align}
where, in an analogy with the usual electrodynamics, we define an effective electric field given by $\mathbfcal{E}=-\textrm{Re}(\xi)\nabla\phi- \partial_t\textbf{A}/\textrm{Re}(\xi)$, while the magnetic field remains unchanged in this scenario, i.e., $\mathbfcal{B}=\nabla\times\textbf{A}$.

On the other hand, from the Bianchi identity,
\begin{align}
\tilde{\pa}_\mu{^{*\!\!}\mathcal{F}_{\mu\nu}}=0, \;\;\;\;\; {^{*\!\!}\mathcal{F}_{\mu\nu}}=\frac12\varepsilon^{\mu\nu\al\be}\mathcal{F}_{\al\be}
\end{align}
there follow the modified Faraday-Lenz (for $\nu= i$) and the Gauss's law for the magnetic field  (for $\nu= 0$):
\begin{align}
& \textrm{Re}(\xi) \nabla \times \mathbfcal{E}=-\pa_t\mathbfcal{B}, \label{Amp_Max} \\
& \textrm{Re}(\xi)\nabla \cdot \mathbfcal{B}=0. 
\end{align}

Following the standard procedure for obtaining the wave equation, let us apply a time derivative on \eqref{Amp_Max}, and use the Equations \eqref{eqMax1} and \eqref{eqMax2}. For non-trivial field configurations, the photonic scattering relation arises from the requirement that 
\begin{equation}\label{rd_foton}
\omega = \pm \textrm{Re}(\xi) |\textbf{k}|.
\end{equation}
For the photino, the field equation is $\ga^{\mu}\tilde{\pa}_\mu\La = 0$, the dispersion relation has the same form as \eqref{rd_foton}.

One of the main consequences of SUSY is the mass degeneracy between a particle and its corresponding superpartner. Since the photon and photino share the same modified dispersion relation, it is immediate to verify that they are both massless. This is simply a confirmation that in the gauge sector there is no SUSY breaking.

\subsection{The modified Dirac and Klein-Gordon equations}
\label{sec4-2}

As we have seen in Section \ref{sec3-3},  due to the fact that the SUSY background is not dynamic, the mixing terms involving the spinors, $\Psi$ and $Z$, also contribute to the extended Dirac action, in addition to the (pure) fermionic sector,
\begin{eqnarray}
S_{\textrm{Dirac}} & = & 
\int d^{4}x\,\bigg\{\dfrac{i}{2}\Big[(\tilde{\pa}_\mu\bar{\Psi})\ga^\mu\Psi-\bar{\Psi}\ga^\mu  \tilde{\pa}_\mu\Psi\Big]+\dfrac{1}{2}v_\mu(\bar{\Psi}\ga^\mu\Psi)-m\bar{\Psi}\Psi
-\frac{1}{16}(\bar{\Psi}\ga^{\mu}\ga_{5}\Psi)(\bar{Z}\ga_{\mu}\ga_{5}Z)
\nn 
\\
&&
-\frac{i}{2}\Big[\bar{Z}\ga^{\mu}(\ph^*\mathcal{P}_{\textrm{L}}-s^{*}\mathcal{P}_{\textrm{R}})\tilde{\pa}_\mu\Psi-(\tilde{\pa}_\mu\bar{\Psi})\ga^{\mu}(\ph\mathcal{P}_{\textrm{L}}-s\mathcal{P}_{\textrm{R}})Z\Big]
-\frac{1}{2}\bar{\Psi}\bigg[\bigg(m\ph-\frac{1}{2}h^{*}s\bigg)\mathcal{P}_{\textrm{L}}
\nn 
\\
&& 
-\bigg(ms+\frac{1}{2}h\ph\bigg)\mathcal{P}_{\textrm{R}}\bigg]Z -\frac{1}{2}\bar{Z}\bigg[\bigg(m\ph^{*}-\frac{1}{2}hs^{*}\bigg)\mathcal{P}_{\textrm{R}}-\bigg(ms^{*}+\frac{1}{2}h^{*}\ph^{*}\bigg)\mathcal{P}_{\textrm{L}}\bigg]\Psi\bigg\}.
\end{eqnarray}
The modified Dirac equation in momentum space is given by
\begin{align}\label{psi0}
\left[(\tilde{p} -\ze)_\mu\ga^\mu + m +R_\mu \gamma^\mu\ga_5 \right]\Psi_0 = & 
-\dfrac{1}{2} \bigg\{ \left[(\tilde{p}_\mu \ga^\mu +m)\varphi_0 -\dfrac{1}{2}h^*s_0 \right]\mathcal{P}_{\textrm{L}} 
\nn 
\\
& 
-\left[(\tilde{p}_\mu \ga^\mu +m)s_0 +\dfrac{1}{2}h\varphi_0 \right]\mathcal{P}_{\textrm{R}} \bigg\}Z,
\end{align}
where $ \ze_{\mu}=\dfrac12v_\mu $ and $ R_{\mu}=\dfrac{1}{16}(\bar{Z}\ga_{\mu}\ga_{5}Z) $. Let us remember that the imaginary part of $\xi$ is a linear function of the coordinates which yields a constant $v_\mu$.

Similarly, the action for the matter scalars reads as below:
\begin{align}
S_{\textrm{KG}} = & 
\,\int d^{4}x\,\bigg\{(\tilde{\pa}^\mu\ph^*)\tilde{\pa}_\mu\ph+(\tilde{\pa}^\mu s^*)\tilde{\pa}_\mu s+\dfrac{i}{2}v_\mu\Big[(\tilde{\pa}^\mu\varphi^*)\ph-\ph^*\tilde{\pa}^\mu\ph+(\tilde{\pa}^\mu s^*)s-s^*\tilde{\pa}^\mu s\Big] 
\nn 
\\
& 
-\bigg(m^{2}+\frac{1}{4}|h|^{2}\bigg)(\ph^{*}\ph+s^{*}s)
-\frac{i}{2}\Big[\bar{Z}\ga^{\mu}(\ph^*\mathcal{P}_{\textrm{L}}-s^{*}\mathcal{P}_{\textrm{R}})\tilde{\pa}_\mu\Psi-(\tilde{\pa}_\mu\bar{\Psi})\ga^{\mu}(\ph\mathcal{P}_{\textrm{L}}-s\mathcal{P}_{\textrm{R}})Z\Big]
\nn 
\\
&
-\frac{1}{2}\bar{\Psi}\bigg[\bigg(m\ph-\frac{1}{2}h^{*}s\bigg)\mathcal{P}_{\textrm{L}}-\bigg(ms+\frac{1}{2}h\ph\bigg)\mathcal{P}_{\textrm{R}}\bigg]Z
-\frac{1}{2}\bar{Z}\bigg[\bigg(m\ph^{*}-\frac{1}{2}hs^{*}\bigg)\mathcal{P}_{\textrm{R}}
\nn 
\\
& 
-\bigg(ms^{*}+\frac{1}{2}h^{*}\ph^{*}\bigg)\mathcal{P}_{\textrm{L}}\bigg]\Psi
\bigg\},
\end{align}
which gives the following modified Klein-Gordon equations:
\begin{align}
\Theta\varphi_0 = &
-\frac{1}{2}\bar{Z}\bigg[\bigg( \tilde{p}_\mu\ga^{\mu}-\frac{1}{2}h^{*}\bigg)\mathcal{P}_{\textrm{L}}+m\mathcal{P}_{\textrm{R}}\bigg]\Psi_0, \label{eomphi2}
\\
\Theta s_0 = &
\,\frac{1}{2}\bar{Z}\bigg[\bigg(\tilde{p}_\mu \ga^{\mu}+\frac{1}{2}h\bigg)\mathcal{P}_{\textrm{R}}
+m\mathcal{P}_{\textrm{L}}\bigg]\Psi_0, \label{eoms2}
\end{align}
where we introduce the useful notation $\Theta \equiv -\bigg[\tilde{p}^2-2(\tilde{p}\cdot \ze)-m^2-\dfrac{1}{4}|h|^2\bigg]$.

As expected from our analysis of the mixing part \eqref{s_interação}, the modified field equations of the matter sector are coupled by means of the background fermion, $Z$. In the next Section, we shall present a way to rewrite these equations and their associated dispersion relations.

\subsection{The matter sector and the associated dispersion relations}
\label{sec4-3}

To obtain the fermionic dispersion relation, let us first substitute the expressions for $\ph_0$ and $s_0$, i.e., \eqref{eomphi2} and \eqref{eoms2}, into the modified Dirac equation \eqref{psi0}. The Fierz rearrangement,
\begin{align}
\left(\bar{Z}M\Psi_0\right)NZ & =-\frac{1}{4}\left[\left(\bar{Z}Z\right)NM+\left(\bar{Z}\gamma_{5}Z\right)N\gamma_{5}M+\left(\bar{Z}\gamma_{\alpha}\gamma_{5}Z\right)N\gamma^{\alpha}\gamma_{5}M\right]\Psi_0,
\end{align}
where $M$ and $N$ are arbitrary 4×4 matrices, allows us to rewrite the pieces involving the mixing of $\Psi_{0}$ and $Z$ in terms of background fermionic condensates, which we will denote just as $\th=\bar{Z}Z$, $\ta=\bar{Z}\ga_{5}Z$ and $C_\al=\bar{Z}\ga_{\al}\ga_{5}Z$. After some simplifications using the relations, $\ga^{\mu}\ga^{\al}\ga^{\nu}=\eta^{\mu\al}\ga^{\nu}+\eta^{\al\nu}\ga^{\mu}-\eta^{\nu\mu}\ga^{\al}-i\ep^{\mu\al\nu\be}\ga_{\be}\ga_{5}$
and $\Si^{\mu\al}=\dfrac{i}{4}\big[\ga^{\mu},\,\ga^{\al}\big]$, the field equation for $\Psi_0$ can be written in compact form as
\begin{align}\label{Dirac_eq}
\Big[(\tilde{p}-\ze)_{\mu}\ga^{\mu}+m-\mathcal{R}_{\mu}\ga^{\mu}\ga_{5}-W_{\mu\al}\Si^{\mu\al}\ga_{5}\Big]\Psi_0 & =0,
\end{align}
where we define the nonlocal structures
\begin{align}
& \mathcal{R}_{\mu} =
\frac{1}{16\Th}\bigg[2\tilde{p}_{\mu}\tilde{p}_{\nu}C^{\nu}
-2\bigg(\tilde{p}\cdot \ze+m^{2}+\frac14|h|^{2}\bigg)C_\mu
+\tilde{p}_\mu(h^*-h)\ta
-\frac12\tilde{p}_\mu(h^*+h)\th\bigg], 
\nn
\\
& W_{\mu\al} =
\frac{im}{4\Th}\tilde{p}_{\mu}C_\al.
\end{align}

The derivation of dispersion relations for the matter sector is a more involved task. In the fermionic case, in particular, the presence of the last term, $\Si^{\mu\al}$, in Equation\,\eqref{Dirac_eq} makes the structure of the Dirac-equation operator more complicated to manipulate. Following the approach presented in ref. \cite{Coll97}, we adopt the standard squaring procedure and apply the Dirac operator with an opposite mass sign to the \eqref{Dirac_eq}:
\begin{equation}\label{sp1}
\mathcal{O}_{2\textrm{-der}}\Psi_0 = 0,
\end{equation}
where the operator is given by
\begin{align}\label{Op}
\mathcal{O}_{2\textrm{-der}} = &
\;(\tilde{p}-\ze)^{2}-m^{2}-\mathcal{R}^{2}
+\dfrac{1}{4}W_{\nu\be}W_{\mu\al}\big[
\left(\eta^{\nu\mu}\eta^{\be\al}-\eta^{\nu\al}\eta^{\be\mu}\right)
+2i(\eta^{\nu\al}\Si^{\be\mu}+\eta^{\be\mu}\Si^{\nu\al})\big]
\nn
\\
&
-i(\tilde{p}-\ze-\mathcal{R})_{\nu}
W_{\mu\al}(\eta^{\nu\mu}\ga^{\al}-\eta^{\nu\al}\ga^{\mu}).
\end{align}
To eliminate the off-diagonal terms in spinor space, we repeat the squaring procedure to Equation\,\eqref{sp1}. Now, let us consider the operator $\mathcal{O}_{2\textrm{-der}}$ with opposite sign for the off-diagonal piece, i.e., $+i(\tilde{p}-\ze-\mathcal{R})_{\ka}W_{\la\rh}(\eta^{\ka\la}\ga^{\rh}-\eta^{\ka\rh}\ga^{\la})$. This procedure gives a fourth-order equation satisfied by each component of the Dirac spinor $\Psi_0$, such that (for non-trivial field configurations)
\begin{align}\label{DRPpsi}
0 = 
&
\;\tilde{p}^{4}
-2\tilde{p}^{2}\bigg[4(\tilde{p}\cdot\ze)
+m^{2}
-\ze^{2}
+\frac14|h|^{2}\bigg]
+8(\tilde{p}\cdot\ze)\big[3(\tilde{p}\cdot\ze)
+m^{2}\big]
\nn
\\
&
+m^{2}\bigg(m^{2}
-2\ze^{2}
+\frac12|h|^{2}\bigg),
\end{align}
where we neglect higher-order terms in the Lorentz-breaking parameters, since it is reasonable to consider them to be small (see, e.g., in refs. \cite{KosPont95,Coll97} for theoretical discussions). The result \eqref{DRPpsi} is the most general expression up to second order in Lorentz breaking for the fermionic dispersion relation of model \eqref{sA1_onShp}.
It is clear its non-dependence on the background fermionic condensates $\theta$, $\tau$ and $C_\al$. Therefore, we can see that, up to second order in LSV, the explicit dependence on the background is manifested only through the auxiliary complex scalar $h$ and vector $\zeta_\mu = \dfrac12 v_\mu$ fields. 
An interesting consequence of this is that the results obtained from \eqref{DRPpsi} are automatically parity invariant.

In the case of scalar fields, $\ph_0$ and $s_0$, the right-hand side of their respective Equations, \eqref{eomphi2} and \eqref{eoms2}, involves a mixing of Dirac and background fermions. To express these equations in terms of the background fermionic condensates, $\th$, $\ta$ and $C_\al$, we first employ the squaring procedure to Equation\,\eqref{psi0}. This allows us to compose an expression for $\Psi_0$, which can then be substituted into the modified Klein-Gordon equations.
After some algebra, we get
\begin{align}
\Theta\ph_0 = &
\,\frac{1}{4\Xi}\big[X^{(1)}(\ph_0,s_0)\th+\mathcal{M}^{(1)}(\ph_0,s_0)\ta+Y^{(1)}_{\al}(\ph_0,s_0)C^{\al}\big], \label{fep}
\\
\Theta s_0 = & 
-\frac{1}{4\Xi}\big[X^{(2)}(\ph_0,s_0)\th+\mathcal{M}^{(2)}(\ph_0,s_0)\ta
+Y^{(2)}_{\al}(\ph_0,s_0)C^{\alpha}\big], \label{fes}
\end{align}
where $\Xi=\big[(\tilde{p}-\ze)^{2}-m^{2}-R^{2}\big]^{2}+4(\tilde{p}-\ze)^{2}R^{2}-4\big[(\tilde{p}-\ze)\cdot R\big]^{2}$, and the explicit form of $X$'s, $\mathcal{M}$'s and $Y_{\al}$'s can be found in Appendix \ref{secB}. 
Finally, from \eqref{fep} and \eqref{fes}, is possible to obtain the full dispersion relation for the scalars, up to second order in LSV,
\begin{align} \label{DRPscalar}
0 = 
&
\;\tilde{p}^{4}
-4\tilde{p}^{2}\bigg[
3(\tilde{p}\cdot \ze)
+\frac{1}{2}m^{2}
-\ze^{2}
+\frac18|h|^{2}\bigg]
+60(\tilde{p}\cdot \ze)^{2}
\nn
\\
&
+4m^{2}\bigg[3(\tilde{p}\cdot \ze)
+\dfrac{1}{32}(\tilde{p} \cdot C)
+\frac{1}{4}m^{2}
-\ze^{2}
+\frac18|h|^{2}\bigg].
\end{align}
In contrast to the fermionic case, Equation\,\eqref{DRPpsi}, the modified dispersion relation for the scalar sector has a non-trivial dependence on the background fermionic condensate $C_\alpha$, in addition to terms proportional to $h$ and $v_\mu$ components.

Now, we are in a position to exploit the effects of this LSV background on the mass spectrum of the fermionic and scalar matter sectors. Let us take the rest frame, where the linear momentum is zero, $\bm{p}=0$, and the rest energy is nothing but the rest mass. From \eqref{DRPpsi}, we verify that the fermionic rest energy results in
\begin{align} \label{fermionMass}
&
E_{0,\,\textrm{fermion}} =
\pm \sqrt{m^2
+\frac12\big(\bm{v}^2
+|h|^2\big)}.
\end{align}
In this scenario, the fermion in question does not allow the existence of tachyonic mass modes. Therefore, considering that LSV terms are very small compared to the mass scale of the particles, we can take the limit $m^2\rightarrow\infty$, which corresponds to the regime $m^2\gg\bm{v}^2$ and $m^2\gg|h|^2$, in the positive solution of the spectra \eqref{fermionMass} in order to obtain a LSV shift in the fermion mass. The result is
\begin{align}
E_{0,\,\textrm{fermion}}\Big|_{m^2\rightarrow\infty} \approx m \left( 1 + \dfrac{\bm v^2 +|h|^2}{4m^2} \right).
\end{align}

On the other hand, for the scalar sector, Equation\,\eqref{DRPscalar}, let us consider two particular cases, for simplicity: 
$i)$ in the previous Section, we have got that the induced SUSY breaking occurs even when the background auxiliary component is zero. So, assuming $h=0$, we arrive at the  following non-parity invariant results 
\begin{align}
& 
E_{0,\,\textrm{scalar}}^{(1)}\Big|_{h=0} =
\,\dfrac{\sqrt{B}}{2} 
\pm\frac12\sqrt{-B+2(2m^2+\bm{v}^2)
-\dfrac{m^2 C_0}{4\sqrt{B}}}, \label{rdh01}
\\
&
E_{0,\,\textrm{scalar}}^{(2)}\Big|_{h=0} =
-\,\dfrac{\sqrt{B}}{2} 
\pm\frac12\sqrt{-B+2(2m^2+\bm{v}^2)
+\dfrac{m^2 C_0}{4\sqrt{B}}}. \label{rdh02}
\end{align}
In the solutions \eqref{rdh01} and \eqref{rdh02} we defined the notations
\begin{align}
A & = 
G 
+\sqrt{G^2-4\left(
16m^4
+16m^2\bm{v}^2 
+\bm{v}^4\right)^3},
\\
G & = 
m^4\bigg[\dfrac{27}{64}\,C_0^2
+64(2\,m^2 
+3\bm{v}^2)\bigg] 
+2\bm{v}^4(30\,m^2
-\bm{v}^2),
\\
B & = 
\dfrac{1}{3}\left\{
\left(\dfrac{A}{2}\right)^{\!\! -\frac{1}{3}}
\left[\left(\dfrac{A}{2}\right)^{\!\! \frac{2}{3}} 
+16m^4 
+16m^2\bm{v}^2
+\bm{v}^4\right] 
+2(2m^2+\bm{v}^2)\right\},
\end{align}
and $C_0=\bar{Z}\ga_0\ga_5Z$ is the zeroth component of the background fermionic condensate, responsible for the parity breaking in both \eqref{rdh01} and \eqref{rdh02} rest mass modes.

$ii)$ In other situation, we assume that the background is invariant under parity transformations. In this case, the effects of the background fermionic condensate disappear, i.e., $C_\al=0$,
\begin{align}
\label{reh0}
E_{0,\,\textrm{scalar}}\Big|_{C_{\alpha}=0} = 
\pm\,\sqrt{m^2
+\bm{v}^2
+\dfrac{1}{2}|h|^2}.
\end{align}
In this situation, the scalar mass spectrum is quite similar to the fermionic case and does not allow for the presence of tachyonic mass modes. As before, assuming that $m^2\gg\bm{v}^2$ and $m^2\gg|h|^2$, \eqref{reh0} reduces to
\begin{align}
E_{0,\,\textrm{scalar}}\Big|_{C_{\alpha}=0} \approx 
m\left(1+
\dfrac{2\bm v^2
+|h|^2}{4m^2}\right).
\end{align}

Unlike from the photon-photino sector, the fermion-scalar sector exhibits a SUSY breaking due the difference between the scalar and fermion rest masses in both cases, parity invariant and parity breaking LSV backgrounds.

Furthermore, another interesting result which can be expressed from the modified dispersion relations is the group velocity,
\begin{align}
\bm{v}_g = 
\bm{\nabla}_{\textrm{Re}(\xi)\bm{p}}E,
\end{align}
where $\bm{\nabla}_{\textrm{Re}(\xi)\bm p} $ is a gradient of linear momentum $\textrm{Re}(\xi)\bm{p}$ and $E$ is the energy that comes from the zeroth component of $\tilde{p}_\mu$. According to the conditions $i)$ and $ii)$, we arrive at the following group velocity  for the scalars
\begin{align}\label{gv_s1}
v^{i}_{g,\,\textrm{scalar}}\Big|_{h=0} = 
& 
\,\frac{\tilde{p}^{\,i}\big[\tilde{\bm p}^{\,2}
+m^2
-E^2
-3(\tilde{\bm p}\cdot\bm v)
+\bm v^2/2\big]}{E\big[\tilde{\bm p}^{\,2}
+m^2
-E^2
-3(\tilde{\bm p}\cdot\bm v)
+\bm v^2/2\big]-m^2C_0/32}
\nn
\\
&
-\frac{3v^i\big[\tilde{\bm p}^{\,2}
+m^2
-E^2
-5(\tilde{\bm p}\cdot\bm v)\big]+m^2C^i/16}
{2E\big[\tilde{\bm p}^{\,2}
+m^2
-E^2
-3(\tilde{\bm p}\cdot\bm v)
+\bm v^2/2\big]-m^2C_0/16},
\end{align}
in the case of $h=0$, and
\begin{align}\label{gv_s2}
v^{i}_{g,\,\textrm{scalar}}\Big|_{C_{\alpha}=0}= 
&  
\,\frac{\tilde{p}^{\,i}}{E}
-\frac{3v^i\big[\tilde{\bm p}^{\,2}
+m^2
-E^2
-5(\tilde{\bm p}\cdot\bm v)\big]}
{2E\big[\tilde{\bm p}^{\,2}
+m^2
-E^2
-3(\tilde{\bm p}\cdot\bm v)
+(2\bm v^2
+|h|^2)/4\big]},
\end{align}
for a parity invariant background.

And finally, the group velocity for the fermionic case is given by
\begin{align}\label{gv_f}
v^i_{g,\,\textrm{fermion}} = 
&
\,\frac{\tilde{p}^{\,i}}{E}
-\frac{v^i\big[\tilde{\bm p}^{\,2}
+m^2
-E^2
-3(\tilde{\bm p}\cdot\bm v)\big]}
{E\big[\tilde{\bm p}^{\,2}
+m^2
-E^2
-2(\tilde{\bm p}\cdot\bm v)
+(\bm v^2
+|h|^2)/4\big]},
\end{align}
where $\tilde{p}_i=\textrm{Re}(\xi)p_i$, with $i=1,2,3$. Both cases reduce to the usual situation, $v^i_g = p^i/E$, if we turn off the LSV background. 
Let us notice that, even in the presence of LSV terms, the expressions for the group velocities \eqref{gv_s1}, \eqref{gv_s2}, and \eqref{gv_f} admit the condition $|v^i_g| < 1$, indicating in this case that the causal structure of the theory is maintained (see, e.g., in refs. \cite{Coll97,AdaKli01,CausUni03,Helayel04} for discussions on causality in other scenarios with Lorentz violation).

\section{The electron's EDM}
\label{sec5}

We are, at this point, ready to start off the task of inspecting the structure of the fermionic current, as the modifications that arise due to the background may induce a contribution to the charged particle  EDM interaction (see, e.g., in refs. \cite{LDM_book,EDM14_Japan} for the textbook level introduction).
First, due to non-local structure, it proves useful to rewrite the modified Equation \eqref{Dirac_eq} as
\begin{align}
\label{D3}
\Big\{(1-s)\big[
(\tilde{p}-\ze)_{\mu}\ga^{\mu}
+m\big]
-r_\mu\ga^\mu\ga_5
-w_{\mu\al}\Si^{\mu\al}\Big\}\Psi_0
&
= 0,
\end{align}
where we introduce the following definitions
\begin{align}
& 
r_{\mu} 
= \frac{1}{16m^2}\bigg[
2\tilde{p}_{\mu}\tilde{p}_{\nu}C^{\nu}
-2\bigg(\tilde{p}\cdot \ze+m^{2}+\frac14|h|^{2}\bigg)C_\mu
+\tilde{p}_\mu(h^*-h)\ta
-\frac12\tilde{p}_\mu(h^*+h)\th\bigg], 
\nn
\\
& 
w_{\mu\al} 
= \frac{i}{4m}\tilde{p}_{\mu}C_\al, 
\nn 
\\
& 
s
= \dfrac{1}{m^2}\big[\tilde{p}^2-2(\tilde{p}\cdot \ze)\big].
\end{align}
By iterating modified Equation \eqref{D3} and taking the Dirac conjugate, we get the Gordon decomposition below
\begin{align}
\bar{\Psi}_0(\tilde{p}')\ga^\nu\Psi_0(\tilde{p}) 
= &
\,-\dfrac{1}{2m} \bar{\Psi}_0(\tilde{p}') \bigg\{\Big[ (\tilde{p}
+\tilde{p}'-2\zeta)^\nu - s(\tilde{p}-\zeta)^\nu 
-s'(\tilde{p}'-\zeta)^\nu\Big]
\nn 
\\
&
\,+\dfrac{i}{2}\left[{\left(w'+w \right)_\mu}^\nu
-{\left(w'+w\right)^\nu}_\mu +2i m (s+s') \delta_\mu^\nu \right]\ga^\mu 
\nn 
\\
&
\,-2i\Big[ (\tilde{p}' - \tilde{p})_\mu  +s(p-\ze)_\mu -  s'(p'-\ze)_\mu\Big] \Si^{\mu\nu}+(r'-r)^\nu \ga_5 
\nn 
\\
&
\, 
+\dfrac{1}{2}\left(w'-w\right)_{\mu\al} \ep^{\mu\al\nu\ta}
\ga_\ta\ga_5\Psi_0(\tilde{p})
-2i(r'+r)_\mu 
(\tilde{p}')\Si^{\mu\nu} \gamma_5  \bigg\} \Psi_0(\tilde{p})
\,.
\label{DCG1}
\end{align}
Here, the primes denote the structures carrying the momentum $\tilde{p}'$.
This result, besides providing information on the vector and axial charge densities in terms proportional to the identity and $\gamma_5$, also brings to light the possible interactions that the electromagnetic current may induce in this scenario, i.e., interactions such as electric and magnetic dipole moments associated with the terms $\Sigma^{\mu\nu}$ and $\Sigma^{\mu\nu}\gamma_5$, respectively, as well as other terms that require further attention.

From the last term of \eqref{DCG1}, one can identify the piece from which the EDM emerges,
\begin{align}
\label{EDM_term}
&
\dfrac{i}{8m^3}C^\ta(p'_\mu p'_\ta+p_\mu p_\ta)\bar{\Psi}_0
(\tilde{p}')\Sigma^{\mu\nu}\ga_5 \Psi_0 (\tilde{p}).
\end{align} 

When minimally coupled to the gauge field $\tilde{a}_\nu$, the term in Equation\,\eqref{EDM_term} corresponds to the following structure for the effective electric dipole interaction
\begin{align}
\mathcal{L}_{\textrm{EDM}} =
-i\,\dfrac{\mu_{\textrm{B}}}{8m^2}\, 
C^{\ta}\tilde{\pa}_{\ta}\big(
\bar{\Psi}\Sigma^{\mu\nu}\ga_5\Psi\big) 
\mathcal{F}_{\mu\nu},
\end{align}
where $\mu_{\textrm{B}}=e/(2m)$ is the Bohr magneton and
$\mathcal{F}_{\mu\nu}$ is defined in Equation \eqref{FF}. Following ref. \cite{EDM91}, one can verify that, in our case, the EDM is given by 
\begin{align}
d = -\,\dfrac{F_3(\tilde{q}^2=0)}{2m}
= \dfrac{\mu_{\textrm{B}}}{4m^2}
C^{\ta}\tilde{q}_{\ta},  
\end{align}
with $\tilde{q} = \tilde{p}' - \tilde{p}$ being the transfer momentum of the interaction.
$F_3(\tilde{q}^2)$ is the third electromagnetic form factor associated with the EDM.

In the case of the electron EDM, denoted by $d_e$, the most recent experimental results, obtained by using electrons confined inside molecular ions subject to a huge intramolecular electric field,
provide an upper bound of $d_e < 4.1 \times 10^{-30}$ e$\cdot$cm \cite{ACME18,JILA23} 
(see also ref. \cite{Yamaguchi20} for the Standard Model prediction concerning the long-distance hadronic contributions).
We can use these results to estimate a bound for the modulus of the fermionic background condensate $C^\ta$.

Let us remember that $\tilde{q}^2=\tilde{q}^{\mu} \tilde{q}_{\mu}=0$ for the EDM form factor configuration. Consequently, assuming a typical energy scale of the order of $1$ MeV and considering the special case where $C^\ta$ is timelike, i.e., $\bm{C}=0$, we arrive at
\begin{align}\label{edm_t}
|C^0| < 2.2 \times 10^{-13}\,\textrm{eV}.
\end{align}
For a purely spacelike $C^\tau$, a reference frame can be chosen in which $C^0=0$. In this case, it is possible to establish a bound on the magnitude of $\bm C$ component,
\begin{align}\label{edm_s}
|\bm{C}| <  \frac{1}{\textrm{Re}(\xi)\cdot\cos(\al)}\,2.2 \times 10^{-13}\,\textrm{eV},
\end{align}
where $\al$ is the angle between the spatial part of the four-momentum transfer, $\tilde{\bm{q}}$, and $\bm C$.
We justify that the choice of splitting the two cases of $C^\ta$ time- or space-like is a common procedure in the study of LSV models. The reason is that either choice (time- or space-like) may induce different effects
as far as physical observables are concerned.

\begin{figure}[H]
\centering
\includegraphics[scale=1]{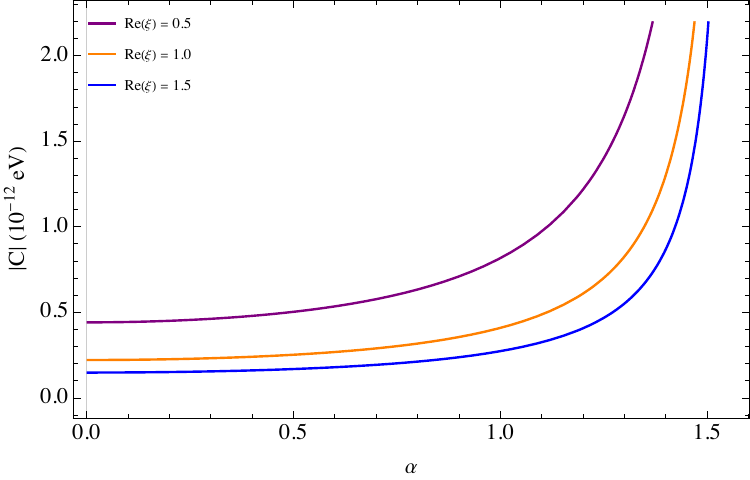}
\begin{quotation}
\vspace{-6mm}
\caption{\small
Plot of the upper bounds on the spatial components of $C^\ta$ as a function of $\al$ for $\textrm{Re}(\xi)=0.5, \; 1.0$, and $1.5$, in purple, orange, and blue, respectively.}
\label{edm}
\end{quotation}
\end{figure}

Once the bound in \eqref{edm_s} depends on both $\textrm{Re}(\xi)$ and $\al$, we can collect all upper bounds that within a range $\al= \big[0,\pi/2\big)$ by plotting a graph of $|\bm C|$ as a function of $\al$ for a specific value of $\textrm{Re}(\xi)$.
As an example, Figure\,\ref{edm} represents the plot of the upper bounds on $|\bm C|$ for $\textrm{Re}(\xi)=0.5, \; 1.0$, and $1.5$.

\section{Conclusive Considerations and Perspectives}
\label{sec6}

In this contribution, we have studied the implications of a Lorentz-violating background supermultiplet of an $\mathcal{N}=1$-supersymmetric $U(1)$ gauge model that extends the topological insulator Isobe-Nagaosa's QED model \cite{IsoNag12}. In our endeavor, we consider that the LSV parameters may have their microscopic origin traced back to a background SUSY multiplet, a chiral superfield. Instead of taking the LSV parameters for granted and extending the SUSY algebra, we keep the ordinary SUSY structure and assume the existence of the background mentioned above. Condensates of the fermionic component of the background superfield appear that are in the backstage of the scalar, pseudoscalar, and pseudovector objects that correspond to LSV parameters.

The modifications in the gauge sector that arise as a consequence of LSV in the fermionic matter consist of nothing but a dilation factor in Maxwell's equations. Indeed, the photon-photino partners are influenced only by the real component of the Lorentz-violating parameter and, therefore, we do not observe any physically relevant modifications in this sector, at least initially. On the other hand, the scalar matter is modified in both its dynamic and mass terms. Furthermore, the fermionic sector also receives an additional contribution that is proportional to the background fermionic condensate, $\bar{Z}\ga_{\mu}\ga_{5}Z$. In this case, we believe that such a contribution arises due to the soft SUSY breaking induced by LSV. In this sense, it is interesting to recall that our model successfully reproduces the CPT-odd terms of the SME, with one of these terms appearing naturally as a direct consequence of SUSY breaking.

Considering the dispersion relations, we verify that, in the case of fermions, the terms proportional to the background fermionic condensates appear only as higher-order contributions in the Lorentz violation parameters. Therefore, by assuming that the latter are small compared to the mass $m$, such condensates do not provide relevant contributions to the rest energy nor to the fermionic group velocity. On the other hand, the scalar dispersion relation receives contribution from the background condensate $\bar{Z}\ga_{\mu}\ga_{5}Z$ already in the second order.

Finally, our results from Section \ref{sec5} indicate a possible connection between the EDM and the SUSY breaking induced by LSV. In this scenario, it is shown that SUSY matter scalars play a supportive role in the process of generating the EDM of the charged fermion.

The introduction of additional Lorentz-violating terms  in the gauge sector, which induces significant modifications in the photon-photino sector, is a path to be followed as a step forward. The motivation to do that is to keep track of the interference between the effects of the different LSV terms present in the matter and the gauge sectors.
  
Furthermore, based on the theoretical framework developed here, we aim to explore potential applications to those condensed matter systems that exhibit (low-energy) SUSY and Lorentz symmetry as emergent phenomena in the dynamics of quasiparticle excitations. On the other hand, the modified photonic and electronic dispersion relations including both SUSY and LSV effects may be studied in connection with photon-photon, Breit-Wheeler and Compton scatterings which may be used to impose upper limits on the parameters with the help of astrophysical data, such as measurements of (non-)attenuation of gamma-rays from active galactic nuclei (AGNs), gamma-ray bursts (GRBs) and extra galactic background light (EBL). Also, we keep in mind that QED high-precision tests can also provide estimates and constraints on the parameters of the model.

\section*{Acknowledgments}
W.C.S. is grateful for the financial support from \textit{Conselho Nacional de
Desenvolvimento Científico e Tecnológico} (CNPq - Brazil) for the Postdoctoral Fellowship [PCI grant number 302494/2024-3].
J.P.S.M. expresses his gratitude to CNPq - Brazil [grant number 140186/2022-0] and \textit{Fundação Carlos Chagas Filho de Amparo à Pesquisa do Estado do Rio de Janeiro} (FAPERJ) [grant number E03/203.638/2024] for granting him PhD Fellowship.



\appendix
\section*{Appendices}
\addcontentsline{toc}{section}{Appendices}
\renewcommand{\thesubsection}{\Alph{subsection}}

\subsection{The alternative case of a vector superfield}
\label{secA}

In this Appendix, we show the result of supersymmetrization following the second approach cited in the Section \ref{sec3-3}, in which we define a vector superfield in the form
\begin{align}
V & = C+\th\ze+\bar{\th}\bar{\ze}+(\th\si^\mu \bar{\th})v_\mu+\th^{2}M+\bar{\th}^{2}M^*+\th^{2}\bar{\th}\bar{\rh}+\bar{\th}^{2}\th\rh+\th^{2}\bar{\th}^{2}D.
\end{align}
Here, the terms in \eqref{cond1} comes out from the combination $ V\big(\Phi_{\textrm{L}}e^{eA_{\textrm{WZ}}}\bar{\Phi}_\textrm{L}-\Phi_{\textrm{R}}e^{-eA_{\textrm{WZ}}}\bar{\Phi}_\textrm{R}\big) \big\vert_{\sim\th^{2} \bar{\th}^{2}} $. Then, after some algebra, one can verify that
\begin{align}\label{sA2}
S = & -\int d^{4}x\,\bigg\{\int d^{2}\th d^{2}\bar{\th}\,\bigg[\bar{\Phi}_{\textrm{L}}e^{eA_{\textrm{WZ}}}\Phi_{\textrm{L}}+\bar{\Phi}_{\textrm{R}}e^{-eA_{\textrm{WZ}}}\Phi_{\textrm{R}}+V\big(\bar{\Phi}_{\textrm{L}}e^{eA_{\textrm{WZ}}}\Phi_{\textrm{L}}-\bar{\Phi}_{\textrm{R}}e^{-eA_{\textrm{WZ}}}\Phi_{\textrm{R}}\big)\bigg] \nn \\
& -\frac14\bigg(\int d^2 \th\, W^{2}+c.c.\bigg)-m\bigg(\int d^{2}\th\,\Phi_{\textrm{R}}\Phi_{\textrm{L}}+c.c.\bigg)\bigg\}.
\end{align}
where
\begin{align}
V\Phi_{\textrm{L}}e^{eA_{\textrm{WZ}}}\bar{\Phi}_\textrm{L} \Big\vert_{\sim\th^{2} \bar{\th}^{2}} = &
-\frac12v_{\mu}(\bar{\psi}\bar{\si}^{\mu}\psi)-\frac{i}{2}v_{\mu}\Big[(\tilde{\pa}^{\mu}\ph^{*})\ph-\ph^{*}\tilde{\pa}^{\mu}\ph\Big]+\ph^{*}\ph D+f^{*}\ph M+f\ph^{*} M^{*} 
\nn 
\\
& -\frac{1}{\sqrt{2}}\Big[\ph\bar{\psi}\bar{\rh}+\ph^{*}\psi\rh+f^{*}\psi\ze+f\bar{\psi}\bar{\ze}\Big]+\frac{i}{2\sqrt{2}}\Big[\ph(\tilde{\pa}_\mu\bar{\psi})\bar{\si}^{\mu}\ze+(\bar{\ze}\bar{\si}^{\mu}\psi)\tilde{\pa}_{\mu}\ph^{*} 
\nn 
\\
& -(\bar{\psi}\bar{\si}^{\mu}\ze)\tilde{\pa}_{\mu}\ph-\ph^{*}\,\bar{\ze}\bar{\si}^{\mu}\tilde{\pa}_\mu\psi\Big]+C\Big[f^{*}f+(\tilde{\pa}^{\mu}\ph^{*})\tilde{\pa}_{\mu}\ph\Big]-\frac{i}{2}C\Big[\bar{\psi}\bar{\si}^{\mu}\tilde{\pa}_\mu\psi 
\nn 
\\
& -(\tilde{\pa}_\mu\bar{\psi})\bar{\si}^{\mu}\psi\Big]+e\bigg\{\frac{iC}{\sqrt{2}}\Big[\ph^{*}(\la\psi)-\ph(\bar{\la}\bar{\psi})\Big]-\frac12C\Big[i(\ph\,\tilde{\pa}^{\mu}\ph^{*}-\ph^{*}\tilde{\pa}^{\mu}\ph) 
\nn 
\\
& +\bar{\psi}\bar{\si}^{\mu}\psi\Big]\tilde{a}_{\mu}-\frac{1}{2\sqrt{2}}\Big[\ph(\bar{\psi}\bar{\si}^{\mu}\ze)+\ph^{*}(\bar{\ze}\bar{\si}^{\mu}\psi)\Big]\tilde{a}_{\mu}+\frac12\ph^{*}\ph\,\bigg[i(\la\ze-\bar{\la}\bar{\ze}) 
\nn 
\\
& +v^{\mu}\tilde{a}_{\mu}+C\bigg(K+\frac{e}{2}\tilde{a}^{2}\bigg)\bigg]\bigg\},
\end{align}
and right sector contribution is
\begin{align}
V\Phi_{\textrm{R}}e^{-eA_{\textrm{WZ}}}\bar{\Phi}_\textrm{R} \Big\vert_{\sim\th^{2} \bar{\th}^{2}} = &
-\frac12v_{\mu}(\bar{\ch}\bar{\si}^{\mu}\ch)+\frac{i}{2}v_{\mu}\Big[(\tilde{\pa}^{\mu}s^{*})s-s^{*}\tilde{\pa}^{\mu}s\Big]+s^{*}s D+gs^{*} M+g^{*}sM^{*} 
\nn 
\\
& -\frac{1}{\sqrt{2}}\Big[s^{*}\bar{\chi}\bar{\rh}+s\chi\rh+g\chi\ze+g^{*}\bar{\chi}\bar{\ze}\Big]+\frac{i}{2\sqrt{2}}\Big[s^{*}(\tilde{\pa}_\mu\bar{\ch})\bar{\si}^{\mu}\ze+(\bar{\ze}\bar{\si}^{\mu}\ch)\tilde{\pa}_{\mu}s 
\nn 
\\
& -(\bar{\ch}\bar{\si}^{\mu}\ze)\tilde{\pa}_{\mu}s^{*}-s\,\bar{\ze}\bar{\si}^{\mu}\tilde{\pa}_\mu\ch\Big]+C\Big[g^{*}g+(\tilde{\pa}^{\mu}s^{*})\tilde{\pa}_{\mu}s\Big]-\frac{i}{2}C\Big[\bar{\ch}\bar{\si}^{\mu}\tilde{\pa}_\mu\chi 
\nn 
\\
& -(\tilde{\pa}_\mu\bar{\chi})\bar{\si}^{\mu}\ch\Big]-e\bigg\{\frac{iC}{\sqrt{2}}\Big[s(\la\ch)-s^{*}(\bar{\la}\bar{\ch})\Big]-\frac12C\Big[i(s^{*}\tilde{\pa}^{\mu}s-s\,\tilde{\pa}^{\mu}s^{*}) 
\nn 
\\
& +\bar{\ch}\bar{\si}^{\mu}\ch\Big]\tilde{a}_{\mu}-\frac{1}{2\sqrt{2}}\Big[s^{*}(\bar{\ch}\bar{\si}^{\mu}\ze)+s(\bar{\ze}\bar{\si}^{\mu}\ch)\Big]\tilde{a}_{\mu}+\frac12s^{*}s\,\bigg[i(\la\ze-\bar{\la}\bar{\ze}) 
\nn 
\\
& +v^{\mu}\tilde{a}_{\mu}+C\bigg(K-\frac{e}{2}\tilde{a}^{2}\bigg)\bigg]\bigg\}.
\end{align}
In this case, using the constraints obtained from the field equations for the auxiliary fields, namely, $ f,\,g $, and $ D $, we get
\begin{align}
S_2 = & 
\int d^{4}x\,\bigg\{-\frac14\mathcal{F}^{2}_{\mu\nu}+\frac{i}{2}\Big[(\tilde{\pa}_\mu\bar{\la})\bar{\si}^{\mu}\la-\bar{\la}\bar{\si}^{\mu}\tilde{\pa}_\mu\la\Big]+\dfrac{i}{2}(1+C)\Big[\bar{\psi}\bar{\si}^\mu(\tilde{\pa}_\mu\psi)-(\tilde{\pa}_\mu\bar{\psi})\bar{\si}^\mu\psi\Big] \nn \\
& +\dfrac{i}{2}(1-C)\Big[\bar{\chi}\bar{\si}^\mu(\tilde{\partial}_\mu\chi)-(\tilde{\partial}_\mu\bar{\chi})\bar{\si}^\mu\chi\Big]+\dfrac{1}{2}v_\mu(\bar{\psi}\bar{\si}^\mu\psi-\bar{\chi}\bar{\si}^\mu\chi)-m(\bar{\psi}\bar{\chi}+\chi\psi)-(1+C) 
\nn 
\\
& \times (\tilde{\pa}^{\mu}\ph^{*})\tilde{\pa}_{\mu}\ph-(1-C)(\tilde{\pa}^{\mu}s^{*})\tilde{\pa}_{\mu}s+\frac{i}{2}v_\mu\Big[(\tilde{\pa}^{\mu}\ph^{*})\ph-\ph^{*}\tilde{\pa}^{\mu}\ph+(\tilde{\pa}^{\mu}s^{*})s-s^{*}\tilde{\pa}^{\mu}s\Big] 
\nn 
\\
& +\bigg[\frac{m^2}{(1-C)}+\frac{M^{*}M}{(1+C)}\bigg]\ph^*\ph+\bigg[\frac{m^2}{(1+C)}+\frac{M^{*}M}{(1-C)}\bigg]s^*s-\dfrac{m}{\sqrt{2}}\bigg[\frac{1}{(1-C)}(\ph\chi\ze+\ph^*\bar{\chi}\bar{\ze}) 
\nn 
\\
& -\frac{1}{(1+C)}(s\bar{\psi}\bar{\ze}+s^*\psi\ze)\bigg]-\frac{i}{2\sqrt{2}}\Big[\ph(\tilde{\pa}_\mu\bar{\psi})\bar{\si}^{\mu}\ze-(\bar{\psi}\bar{\si}^\mu\ze)\tilde{\pa}_\mu\ph-\ph^*\bar{\ze}\bar{\si}^\mu\tilde{\pa}_\mu\psi+(\bar{\ze}\bar{\si}^\mu\psi)\tilde{\pa}_\mu\ph^* 
\nn 
\\
& +s\bar{\ze}\bar{\si}^\mu\tilde{\pa}_\mu\ch-(\bar{\ze}\bar{\si}^\mu\ch)\tilde{\pa}_{\mu}s-s^{*}(\tilde{\pa}_\mu\bar{\ch})\bar{\si}^{\mu}\ze+(\bar{\ch}\bar{\si}^\mu\ze)\tilde{\pa}_{\mu}s^*\Big]+\frac{1}{\sqrt{2}}\Big[(\ph\bar{\psi}-s^{*}\bar{\ch})\bar{\rh} 
\nn 
\\
& +(\ph^{*}\psi-s\ch)\rh\Big]+\dfrac{1}{2(1+C)}\Big[\bar{\ze}\bar{\psi}\ze\psi-\sqrt{2}(M\ph\bar{\ze}\bar{\psi}+M^{*}\ph^{*}\ze\psi)\Big]+\dfrac{1}{2(1-C)}\Big[\bar{\ze}\bar{\chi}\ze\chi 
\nn 
\\
& -\sqrt{2}(Ms^{*}\bar{\ze}\bar{\chi}+M^{*}s\ze\chi)\Big]-(\ph^{*}\ph-s^{*}s)D-\frac{e}{2}(\ph^{*}\ph+s^{*}s)\Big[i(\la\ze-\bar{\la}\bar{\ze})+v^{\mu}\tilde{a}_{\mu}\Big] 
\nn 
\\
& +\frac{e}{2\sqrt{2}}(\ph\bar{\psi}\bar{\si}^{\mu}\ze+\ph^{*}\bar{\ze}\bar{\si}^{\mu}\psi+s^{*}\bar{\ch}\bar{\si}^{\mu}\ze+s\bar{\ze}\bar{\si}^{\mu}\ch)\tilde{a}_{\mu}-\frac{e}{2}(1+C)\Big[i\sqrt{2}(\ph^{*}\la\psi-\ph\bar{\la}\bar{\psi}) 
\nn 
\\
& +(\bar{\psi}\bar{\si}^{\mu}\psi)\tilde{a}_{\mu}-i(\ph\tilde{\pa}^{\mu}\ph^{*}-\ph^{*}\tilde{\pa}^{\mu}\ph)\tilde{a}_{\mu}\Big]+\frac{e}{2}(1-C)\Big[i\sqrt{2}(s\la\ch-s^{*}\bar{\la}\bar{\ch}) 
\nn 
\\
& 
+i(s\tilde{\pa}^{\mu}s^{*}-s^{*}\tilde{\pa}^{\mu}s)\tilde{a}_{\mu}
-(\bar{\ch}\bar{\si}^{\mu}\ch)\tilde{a}_{\mu}\Big]-\frac{e^{2}}{4}\Big[(1+C)\ph^{*}\ph+(1-C)s^{*}s\Big]\tilde{a}^{2}
\nn 
\\
&
-\frac{e^{2}}{8}\Big[(1+C)\ph^{*}\ph-(1-C)s^{*}s\Big]^{2} \bigg\}.
\end{align}

As in the first approach, the gauge sector is given by \eqref{s_calibre}. For the other sectors, we have
\begin{itemize}
\item fermionic sector
\begin{align}
S_{\textrm{fermion}} = & 
\int d^{4}x\,\bigg\{\dfrac{i}{2}\Big[\bar{\Psi}\ga^\mu  \tilde{\pa}_\mu\Psi-(\tilde{\pa}_\mu\bar{\Psi}) \ga^\mu\Psi\Big]+\dfrac{1}{2}v_\mu(\bar{\Psi}\ga^\mu\Psi)-m\bar{\Psi}\Psi \nn \\
& +\dfrac{e}{2}\big(\bar{\Psi}\ga^{\mu}\ga_{5}\Psi-C\bar{\Psi}\ga^{\mu}\Psi\big)\tilde{a}_{\mu}-\dfrac{i}{2}C\Big[\bar{\Psi}\ga^\mu\ga_{5} \tilde{\pa}_\mu\Psi-(\tilde{\pa}_\mu\bar{\Psi}) \ga^\mu\ga_{5}\Psi\Big]\bigg\}.
\end{align}	
\item scalar sector
\begin{align}
S_{\textrm{scalar}} = & 
-\int d^{4}x\,\bigg\{(1+C)(\tilde{D}^{\mu}\ph^{*})\tilde{D}_{\mu}\ph+(1-C)(\tilde{D}^{\mu}s^{*})\tilde{D}_{\mu}s+(\ph^{*}\ph-s^{*}s)D \nn \\
& +iv_\mu\big(\ph^{*}\tilde{D}^{\mu}\ph+s^{*}\tilde{D}^{\mu}s\big)-\bigg[\frac{m^2}{(1-C)}+\frac{M^{*}M}{(1+C)}\bigg]\ph^*\ph \nn \\
& -\bigg[\frac{m^2}{(1+C)}+\frac{M^{*}M}{(1-C)}\bigg]s^*s\bigg\}.
\end{align}	
\item gaugino and matter self-interaction sector
\begin{align}
S_{\textrm{int}} = & 
\int d^{4}x\,\bigg\{\frac{ie}{\sqrt{2}}\bigg[(1+C)(\ph\bar{\Psi}\mathcal{P}_{\textrm{R}}\La-\ph^{*}\bar{\La}\mathcal{P}_{\textrm{L}}\Psi)+(1-C)(s\bar{\Psi}\mathcal{P}_{\textrm{L}}\La-s^{*}\bar{\La}\mathcal{P}_{\textrm{R}}\Psi)\bigg] 
\nn 
\\
& +\frac{1}{\sqrt{2}}\Big[\bar{\Psi}(\ph\mathcal{P}_{\textrm{R}}-s\mathcal{P}_{\textrm{L}})R+\bar{R}(\ph^{*}\mathcal{P}_{\textrm{L}}-s^{*}\mathcal{P}_{\textrm{R}})\Psi\Big]-\frac{i}{2\sqrt{2}}\Big[(\tilde{D}^{*}_\mu\bar{\Psi})\ga^{\mu}(\ph\mathcal{P}_{\textrm{L}}-s\mathcal{P}_{\textrm{R}})Z 
\nn 
\\
& -\bar{Z}\ga^\mu(\ph^*\mathcal{P}_{\textrm{L}}-s^{*}\mathcal{P}_{\textrm{R}})\tilde{D}_\mu\Psi-\bar{\Psi}\ga^{\mu}(\tilde{\pa}_\mu\ph\mathcal{P}_{\textrm{L}}-\tilde{\pa}_{\mu}s\mathcal{P}_{\textrm{R}})Z
\nn 
\\
&
+\bar{Z}\ga^\mu(\tilde{\pa}_\mu\ph^*\mathcal{P}_{\textrm{L}}-\tilde{\pa}_{\mu}s^*\mathcal{P}_{\textrm{R}})\Psi\Big]
-\dfrac{m}{\sqrt{2}}\bigg[\frac{1}{(1-C)}(\ph\bar{\Psi}\mathcal{P}_{\textrm{L}}Z+\ph^*\bar{Z}\mathcal{P}_{\textrm{R}}\Psi) 
\nn 
\\
&
-\frac{1}{(1+C)}(s\bar{\Psi}\mathcal{P}_{\textrm{R}}Z+s^*\bar{Z}\mathcal{P}_{\textrm{L}}\Psi)\bigg]
+\dfrac{1}{2(1+C)}\Big[(\bar{\Psi}\mathcal{P}_{\textrm{R}}Z)(\bar{Z}\mathcal{P}_{\textrm{L}}\Psi)
\nn 
\\
&  
-\sqrt{2}(M\ph\bar{\Psi}\mathcal{P}_{\textrm{R}}Z+M^{*}\ph^{*}\bar{Z}\mathcal{P}_{\textrm{L}}\Psi)\Big]
+\dfrac{1}{2(1-C)}\Big[(\bar{Z}\mathcal{P}_{\textrm{R}}\Psi)(\bar{\Psi}\mathcal{P}_{\textrm{L}}Z) 
\nn 
\\
& -\sqrt{2}(Ms^{*}\bar{Z}\mathcal{P}_{\textrm{R}}\Psi+M^{*}s\bar{\Psi}\mathcal{P}_{\textrm{L}}Z)\Big]-\frac{ie}{4}(\ph^{*}\ph+s^{*}s)\big[\bar{\La}\ga_{5}Z+\bar{Z}\ga_{5}\La\big]
\nn 
\\
& -\frac{e^{2}}{8}\Big[(1+C)\ph^{*}\ph-(1-C)s^{*}s\Big]^{2}\bigg\},
\end{align}
where
\begin{align}
R = \begin{pmatrix}
\rh_\al \\ \bar{\rh}^{\dot{\al}}
\end{pmatrix}.
\end{align}
\end{itemize}

\subsection{Useful results}
\label{secB}

We collect here some useful results related to the calculation of dispersion relations in Sect.\,\ref{sec4-3}.
\begin{align}
X^{(1)}(\ph_0,s_0) = &
\,\dfrac{1}{16}\,\bigg\{\Big[
2m^2(\tilde{p}\cdot\ze)
-2(\tilde{p}^2
-\tilde{p}\cdot\mathcal{R} 
-\tilde{p}\cdot\ze)
(\tilde{p}^2
+2\tilde{p}\cdot\mathcal{R}  
-2\tilde{p}\cdot\ze
\nn 
\\
&
+\mathcal{R}^2
-2\ze\cdot\mathcal{R} 
+\ze^2)\Big]h
-2\Big[2m^2\tilde{p}^2 
-3m^2\tilde{p}\cdot\ze 
-\tilde{p}^4
-2(\tilde{p}\cdot\mathcal{R})^2
\nn 
\\
&
-(\tilde{p}\cdot\ze)(2\tilde{p}\cdot\ze 
+3\mathcal{R}^2 
-2\ze\cdot\mathcal{R} 
-\ze^2)
-(\tilde{p}\cdot\mathcal{R})(\mathcal{R}^2 
-2\ze\cdot\mathcal{R}+\ze^2) 
\nn 
\\
& 
+\tilde{p}^2(\tilde{p}\cdot\mathcal{R}
+3\tilde{p}\cdot\ze 
+3\mathcal{R}^2
-2\ze\cdot\mathcal{R}
-\ze^2)
\Big]h^{*}
+2m^{2}\big(m^{2}+\tilde{p}\cdot \mathcal{R}
\nn 
\\
& 
+\mathcal{R}^{2}
-\tilde{\ze}^{2}\big)(h+h^{*})\bigg\}\ph_0
+\dfrac{1}{16}\,\bigg\{
4m^5
+4m^3\big(2\tilde{p}\cdot \tilde{\ze}
+\mathcal{R}^{2}
-\tilde{\ze}^{2}\big)
\nn 
\\
& 
-4m\Big[
\tilde{p}^2\big(\tilde{p}^2
-2\tilde{p}\cdot \tilde{\ze}
+3\mathcal{R}^{2}
+\tilde{\ze}^{2}\big)
-4(\tilde{p}\cdot\mathcal{R})\big(
\tilde{p}\cdot \mathcal{R}
-\ze\cdot\mathcal{R}
\big)
\nn 
\\
& 
-4\tilde{p}\cdot\ze \mathcal{R}^2\Big]       
-m\big(m^2
-\tilde{p}^2
+2\tilde{p}\cdot \tilde{\ze}
+\mathcal{R}^{2}
-\tilde{\ze}^{2}\big)h^{*2}\bigg\}s_0,
\end{align} 

\begin{align}
X^{(2)}(\ph_0,s_0) = &
\,-\dfrac{1}{16}\,\bigg\{
4m^3(m^2
-2\tilde{p}^2
+2\tilde{p}\cdot\mathcal{R} 
+4\tilde{p}\cdot\ze 
+\mathcal{R}^2
-\ze^2)
+4m\Big[\tilde{p}^2(\tilde{p}^2
-2\tilde{p}\cdot\mathcal{R}
\nn 
\\
&
-4\tilde{p}\cdot\ze 
-5\mathcal{R}^2 
+\ze^2)
+2(\tilde{p}\cdot\mathcal{R})
(2\tilde{p}\cdot\mathcal{R}
+2\tilde{p}\cdot\ze 
+\mathcal{R}^2
-2\ze\cdot\mathcal{R}
-\ze^2)
\nn 
\\
&
+2(\tilde{p}\cdot\ze)              
(2\tilde{p}\cdot\ze 
+3\mathcal{R}^2
-\ze^2) 
+m(m^2
-\tilde{p}^2
+2\tilde{p}\cdot\ze 
+\mathcal{R}^2
-\ze^2)h^*h\Big]
\bigg\}\varphi_0
\nn 
\\
&
+\,\dfrac{1}{16}\,\bigg\{4m^2(m^2
-2\tilde{p}^2
+3\tilde{p}\cdot\ze
+\mathcal{R}^2
-\ze^2)
+4\Big[\tilde{p}^2(\tilde{p}^2 
-3\tilde{p}\cdot\ze 
-3\mathcal{R}^2
\nn 
\\
&
-2\ze\cdot\mathcal{R} 
+\ze^2)  
+2(\tilde{p}\cdot\mathcal{R})
(\tilde{p}\cdot\mathcal{R}
+\tilde{p}\cdot\ze 
-\ze\cdot\mathcal{R}
-\ze^2) 
\nn 
\\
&
+(\tilde{p}\cdot\ze)
(2\tilde{p}\cdot\ze 
+3\mathcal{R}^2
+2\ze\cdot\mathcal{R}
-\ze^2)\Big]              
\bigg\}h^* s_0,
\end{align}

\begin{align}
\mathcal{M}^{(1)}(\ph_0,s_0) = &
\,\dfrac{1}{16}\,\bigg\{\Big[ 
2m^2(\tilde{p}\cdot\ze)
-2(\tilde{p}^2 -\tilde{p}\cdot\mathcal{R} 
-\tilde{p}\cdot\ze)
(\tilde{p}^2+2\tilde{p}\cdot\mathcal{R} 
-2\tilde{p}\cdot\ze +\mathcal{R}^2
\nn 
\\
& 
-2\ze\cdot\mathcal{R}+\ze^2) 
\Big]h
+\Big[4m^2\tilde{p}^2 
-6m^2\tilde{p}\cdot\ze 
-2\big[\tilde{p}^4
+2(\tilde{p}\cdot\mathcal{R})^2
\nn 
\\
& 
+(\tilde{p}\cdot\ze)(2\tilde{p}\cdot \ze 
+3\mathcal{R}^2 
-2\ze\cdot\mathcal{R} 
-\ze^2)\big]
+(\tilde{p}\cdot\mathcal{R})(\mathcal{R}^2 
-2\ze\cdot\mathcal{R}
+\ze^2) 
\nn 
\\
& 
+\tilde{p}^2(-\tilde{p}\cdot\mathcal{R}
-3\tilde{p}\cdot\ze 
-3\mathcal{R}^2
+2\ze\cdot\mathcal{R}
+\ze^2)\Big]h^{*}
+2m^{2}\big(m^{2}+\tilde{p}\cdot\mathcal{R}
\nn 
\\
& 
+\mathcal{R}^{2}
-\tilde{\ze}^{2}\big)(h-h^{*})\bigg\}\ph_0
+\dfrac{1}{16}\,\bigg\{4m^5
+4m^3\big(2\tilde{p}\cdot \tilde{\ze}
+\mathcal{R}^{2}
-\tilde{\ze}^{2}\big)
\nn 
\\
& 
-4m\Big[\tilde{p}^2\big(\tilde{p}^2
-2\tilde{p}\cdot \tilde{\ze}
+3\mathcal{R}^{2}
+\tilde{\ze}^{2}\big)
-4(\tilde{p}\cdot\mathcal{R})\big(
\tilde{p}\cdot \mathcal{R}
-\ze\cdot\mathcal{R}
\big)
\nn 
\\
& 
-4(\tilde{p}\cdot\ze)\mathcal{R}^2\Big]       
+m\big(m^2-\tilde{p}^2
+2\tilde{p}\cdot\tilde{\ze}
+\mathcal{R}^{2}
-\tilde{\ze}^{2}\big)h^{*2}\bigg\}s_0,
\end{align} 

\begin{align}
\mathcal{M}^{(2)}(\ph_0,s_0) =  &
\,\dfrac{1}{16}\,\bigg\{4m^3(m^2-2\tilde{p}^2
+2 \tilde{p}\cdot\mathcal{R} 
+4\tilde{p}\cdot\ze 
+\mathcal{R}^2
-\ze^2)
+4m\Big[\tilde{p}^2(\tilde{p}^2
-2\tilde{p}\cdot\mathcal{R}
\nn 
\\
&
-4\tilde{p}\cdot\ze 
-5\mathcal{R}^2 
+\ze^2)
+2(\tilde{p}\cdot\mathcal{R})
(2\tilde{p}\cdot\mathcal{R}
+2\tilde{p}\cdot\ze 
+\mathcal{R}
-2\ze\cdot\mathcal{R}
-\ze^2)
\nn 
\\
&                     
+2(\tilde{p}\cdot\ze)              
(2\tilde{p}\cdot\ze 
+3\mathcal{R}^2
-\ze^2)\Big]
+m(m^2-\tilde{p}^2
+2\tilde{p}\cdot\ze
+\mathcal{R}^2-\ze^2)h^*h 
\bigg\}\varphi_0
\nn 
\\
&
-\,\dfrac{1}{16}\,\Big[
4m^2(\tilde{p}\cdot\mathcal{R})
-4(\tilde{p}\cdot\mathcal{R})(\tilde{p}^2 
-2\tilde{p}\cdot\ze) 
+\mathcal{R}^2
-\ze^2)\Big]      
h^* s_0,
\end{align}

\begin{align}
Y^{(1)}_\al(\ph_0,s_0) = &
\,-\dfrac{1}{16}\, 
\Big[4m\big(m^2
-\tilde{p}^2
+2\tilde{p}\cdot\ze 
-\mathcal{R}^2
-\ze^2\big)\tilde{p}_\al 
+8m\mathcal{R}^2\ze_\al
-4m\big(m^2
-\tilde{p}^2
\nn 
\\
& 
+2\tilde{p}\cdot\mathcal{R}
+2\tilde{p}\cdot\ze 
-\mathcal{R}^2
-\ze^2 \big) \mathcal{R}_\al 
\Big]h^* s_0
+\dfrac{1}{16}\,\bigg\{ 
\Big[4m^2 \big(
m^2
-2\tilde{p}^2
+4\tilde{p}\cdot\ze
\nn 
\\
&  
-\mathcal{R}^2
-2\ze\cdot\mathcal{R}
-\ze^2\big) 
+4\tilde{p}^2 
\big(\tilde{p}^2 
-4\tilde{p}\cdot\mathcal{R}
-4\tilde{p}\cdot\ze 
+\mathcal{R}^2
-2\ze\cdot\mathcal{R}
+\ze^2\big)
\nn 
\\
& 
+8(\tilde{p}\cdot\mathcal{R})
\big(-2\tilde{p}\cdot\mathcal{R}
+4\tilde{p}\cdot\ze 
+\mathcal{R}^2
-\ze^2\big)
+8(\tilde{p}\cdot\ze)
\big(2\tilde{p}\cdot\ze
+\mathcal{R}^2
-\ze^2\big)
\Big] \tilde{p}_\al
\nn 
\\
&   
+\Big[4m^2\big(m^2
-2\tilde{p}^2
+4\tilde{p}\cdot\ze 
+3\mathcal{R}^2
+2\ze\cdot\mathcal{R}
-\ze^2\big)
+4\tilde{p}^2\big(\tilde{p}^2 
+2\tilde{p}\cdot\mathcal{R}
\nn 
\\
& 
-2\tilde{p}\cdot\ze 
-3\mathcal{R}^2
+2\ze\cdot\mathcal{R}
+\ze^2\big)
+8(\tilde{p}\cdot\mathcal{R})
\big(2\tilde{p}\cdot\mathcal{R}
-2\tilde{p}\cdot\ze \big)
\Big]\ze_\al 
\nn 
\\
& 
+4\Big[m^2\big(m^2
+2\tilde{p}\cdot\mathcal{R} 
+2\tilde{p}\cdot\ze 
+\mathcal{R}^2
-2\ze\cdot\mathcal{R}
-3\ze^2\big)
+\tilde{p}^2\big(3\tilde{p}^2   
-2\tilde{p}\cdot\mathcal{R}
\nn 
\\
&
+6\tilde{p}\cdot\ze 
+\mathcal{R}^2
-\ze\cdot\mathcal{R}
-\ze^2\big)
+4(\tilde{p}\cdot\ze) 
\big(\tilde{p}\cdot\ze 
-\tilde{p}\cdot\mathcal{R}\big)
\Big]\mathcal{R}_\al 
\nn 
\\
&  
-16i\big(\tilde{p}^2
-\tilde{p}\cdot\mathcal{R}
-\tilde{p}\cdot\ze\big)
{\varepsilon_\al}^{\mu\nu\kappa}
\tilde{p}_\mu \mathcal{R}_{\nu}\ze_\kappa
\bigg\}\ph_0,
\end{align}

\begin{align}
Y^{(2)}_\al(\ph_0,s_0) = & 
\,\dfrac{1}{16}\, 
\bigg\{ m\Big[4\big(m^2
-\tilde{p}^2
+2\tilde{p}\cdot\mathcal{R}
+2\tilde{p}\cdot\ze
+2\mathcal{R}^2
-\ze\cdot\mathcal{R}
-\ze^2
\big)\tilde{p}_\al
-2\big(m^2
-\tilde{p}^2
\nn 
\\
&
+4\tilde{p}\cdot\mathcal{R}
+2\tilde{p}\cdot\ze
+3\mathcal{R}^2
-2\ze\cdot\mathcal{R}
-\ze^2 \big)\ze_\al
+2\big(m^2
-5\tilde{p}^2
-2\tilde{p}\cdot\mathcal{R}
\nn 
\\
&
+8\tilde{p}\cdot\ze
+\mathcal{R}^2
+\ze\cdot\mathcal{R}
-3\ze^2
\big)\mathcal{R}_\al
-4i{\varepsilon_\al}^{\mu\nu\ka}
\tilde{p}_\mu\mathcal{R}_\nu\ze_\ka\Big]h
-\Big[4\big(\mathcal{R}^2
+\ze\cdot\mathcal{R}
\big)\tilde{p}_\al
\nn 
\\
&              
-2\big(m^2
-\tilde{p}^2
+2\tilde{p}\cdot\ze
+3\mathcal{R}^2
+2\ze\cdot\mathcal{R}
-\ze^2 \big)\ze_\al 
-2\big(m^2
-\tilde{p}^2
+2\tilde{p}\cdot\mathcal{R}
\nn 
\\
&
+4\tilde{p}\cdot\ze
+\mathcal{R}^2
-2\ze\cdot\mathcal{R}
-3\ze^2
\big)\mathcal{R}_\al
-4i{\varepsilon_\al}^{\mu\nu\kappa}
\tilde{p}_\mu\mathcal{R}_\nu\ze_\kappa
\Big]h^*
\bigg\} \ph_0
\nn 
\\
& +
\,\dfrac{1}{16}\, 
\bigg\{\Big[\big(m^2
-\tilde{p}^2
-2\tilde{p}\cdot\mathcal{R}
+2\tilde{p}\cdot\ze+3\mathcal{R}^2
+2\ze\cdot\mathcal{R}
-\ze^2
\big)\tilde{p}_\al
-\big(m^2
-\tilde{p}^2
\nn 
\\
&
-2\tilde{p}\cdot\mathcal{R}
+2\tilde{p}\cdot\ze-3\mathcal{R}^2
+2\ze\cdot\mathcal{R}
-\ze^2 \big)\ze_\al
-\big(m^2
-3\tilde{p}^2
+2\tilde{p}\cdot\mathcal{R}
\nn 
\\
&
+6\tilde{p}\cdot\ze
+\mathcal{R}^2
-2\ze\cdot\mathcal{R}
-3\ze^2
\big)\mathcal{R}_\al
\Big]{h^*}^2
+4\Big[3m^4 
+\tilde{p}^4
+4(\tilde{p}\cdot\mathcal{R})^2
\nn 
\\
&
+4(\tilde{p}\cdot\ze)^2
-2(\tilde{p}\cdot\mathcal{R}-\tilde{p}\cdot\ze)\mathcal{R}^2
+2(\tilde{p}\cdot\mathcal{R}
-\tilde{p}\cdot\ze)\ze^2
-m^2(\tilde{p}^2  
-8\tilde{p}\cdot\ze
\nn 
\\
&                 
-5\mathcal{R}^2
+2\ze\cdot\mathcal{R}
+3\ze^2)
-\tilde{p}^2(4\tilde{p}\cdot\ze
+\mathcal{R}^2 
+2\ze\cdot\mathcal{R}
-\ze^2)
\Big]\tilde{p}_\al
\nn 
\\
& 
-4\Big[m^4 
-\tilde{p}^4
+4(\tilde{p}\cdot\mathcal{R})^2
+4(\tilde{p}\cdot\mathcal{R})(\tilde{p}\cdot\ze)
+m^2(2\tilde{p}\cdot\mathcal{R}
+2\tilde{p}\cdot\ze
+3\mathcal{R}^2
\nn 
\\
&
-2\ze\cdot\mathcal{R}
-\ze^2)
-\tilde{p}^2(2\tilde{p}\cdot\mathcal{R}
-2\tilde{p}\cdot\ze
+\mathcal{R}^2 
+2\ze\cdot\mathcal{R}
+\ze^2)\Big]\ze_\al
\nn 
\\
& 
+4\Big[m^4 
+\tilde{p}^4
+4(\tilde{p}\cdot\ze)(
\tilde{p}\cdot\mathcal{R}
+\tilde{p}\cdot\ze)
-m^2(2\tilde{p}^2  
+2\tilde{p}\cdot\mathcal{R}
-6\tilde{p}\cdot\ze
-\mathcal{R}^2
\nn 
\\
&
-2\ze\cdot\mathcal{R}
+3\ze^2)
-\tilde{p}^2(2\tilde{p}\cdot\mathcal{R}
+2\tilde{p}\cdot\ze
-\mathcal{R}^2 
+2\ze\cdot\mathcal{R}
-\ze^2)
\Big]\mathcal{R}_\al
\nn 
\\
&              
-16i(\tilde{p}^2
-\tilde{p}\cdot\mathcal{R}
-\tilde{p}\cdot\ze)
{\varepsilon_\al}^{\mu\nu\kappa}
\tilde{p}_\mu\mathcal{R}_\nu\ze_\kappa
\bigg\} s_0.
\end{align}



\end{document}